\documentclass[%
amsmath,
reprint,
amsmath,amssymb,
aps,
]{revtex4-1}
\usepackage[T1]{fontenc}
\usepackage{textcomp}
\usepackage{txfonts}
\usepackage{bm}
\usepackage{graphicx}
\usepackage{dcolumn}
\usepackage{bm}
\usepackage{xcolor}
\usepackage{amsmath}
\usepackage{lipsum}
\usepackage{mathtools}
\usepackage{sidecap}
\usepackage{booktabs}
\usepackage{graphicx}
\usepackage{float}
\usepackage{anyfontsize}
\usepackage{hyperref}
\usepackage[utf8]{inputenc}
\usepackage{textgreek}
\usepackage{multirow}

\hypersetup{colorlinks=true, linkcolor=blue, citecolor=blue, urlcolor=blue}

\begin{document}

\preprint{APS/123-QED}

\title{Systematic study of superheavy nuclei within a microscopic collective Hamiltonian: Impact of quantum shape fluctuations}

\author{X. Q. Yang$^{1}$}
\author{R. Y. Hu$^{1}$}
\email{The author contributed equally to this work}
\author{R. N. Mao$^{1}$}
\author{J. Xiang$^{2}$}
\author{Z. P. Li$^{1}$}
\email{E-mail: zpliphy@swu.edu.cn}
\affiliation{%
$^{1}$School of Physical Science and Technology, Southwest University, Chongqing 400715, China \\
$^{2}$College of Physics and Electronic Engineering, Chongqing Normal University, Chongqing 401331, China\\
School of Physics and Electronic, Qiannan Normal University for Nationalities, Duyun 558000, China}
\date{\today}

\begin{abstract}
The even-even superheavy nuclei with $104 \leqslant  Z \leqslant 126$ and  $N\leqslant 258$  have been investigated using a microscopic five-dimensional collective Hamiltonian (5DCH) based on constrained triaxial relativistic Hartree-Bogoliubov calculations with the PC-PK1 density functional. The 5DCH approach effectively captures the characteristic of isospin dependence of nuclear binding energies, two-nucleon separation energies, and $\alpha$-decay energies across isotopic chains and demonstrates consistent accuracy as $Z$ increases, underscoring the model's predictive power. The collective potentials, average quadrupole deformations, and characteristic collective observables: $E(2^+_1)$, $R_{42}$, and $B(E2; 2^+_1\to 0^+_1)$ reveal a shape transition from well-prolate deformation around $N=150$ and $N=210$ to medium-deformed $\gamma$-soft shape around $N=176$ and $N=246$, and finally to a spherical shape near $N=184$ and $N=258$ for the isotopic chains with $104\leqslant Z\leqslant 118$. Oblate deformations are favored for $Z\geqslant 120$ isotopes around $N=178$. Remarkably, for a substantial range of transitional superheavy nuclei with $N\gtrsim184$ and $N\gtrsim240$, no $0^+$ states bounded by the fission saddles are predicted within their very shallow potential wells due to quantum shape fluctuations (QSFs). Additionally, sharp variations predicted for two-neutron separation energies $S_{2n}$ and $\alpha$-decay energies $Q_\alpha$ at $N=184$ and $258$ in mean-field calculations are significantly reduced and shifted to $N=182$ and $256$ in the 5DCH calculations, which is caused by the rapid evolution of the dynamical correlation energies related to QSFs around the nuclear spherical  shells.
\end{abstract}

\maketitle
\section{Introduction}
Over the past two decades, the nuclear chart has been significantly expanded at the superheavy region with the synthesis of new elements up to $Z=118$ \cite{Hofmann2000RMP,Hamilton2013ARNPS,Hofmann2016EPJA,Oganessian2015RPP,Oganessian2017PS,Morita2015NPA,Smits2024NRP}. The experimental investigation of superheavy nuclei (SHN) poses substantial challenges, and considerable resources are currently devoted to this frontier of nuclear physics research \cite{Hofmann2009PPNP,Kaur2022NPA,Giuliani2019RMP,Oganessian2012PRL,Munzenberg2015NPA,Oganessian2015NPA,Ackermann2024EPJA,Munzenberg2023EPJA,Lommel2023EPJA,Gates2022EPJA,Zhou2020SSPMA,Heinz2022EPJA,Adamian2022EPJA,Lopez-Martens2022EPJA,Gan2022EPJA,Itkis2022EPJA}. Advanced experimental techniques have begun to provide initial insights into nuclear structure beyond basic characteristics such as $\alpha$-decay half-lives and $Q_\alpha$ values \cite{Rudolph2013PRL,Forsberg2016NPA}. Additionally, the chemical properties of some of these new elements have been explored \cite{Yakushev2014IC}. Looking ahead, dedicated facilities in Dubna \cite{Dmitriev2016EPJ,Adamian2022EPJA,Oganessian2022NIMPRSA,Lopez-Martens2022EPJA} and RIKEN \cite{Sakai2022EPJA} are expected to greatly enhance the production of SHN for both physics and chemistry. Other significant contributors to SHN research include GSI/FAIR in Germany \cite{Munzenberg2015NPA,Hofmann2016EPJA,Munzenberg2023EPJA}, ORNL \cite{Ackermann2019APPB} and Berkeley \cite{Gates2018PRL,Gates2022EPJA,GatesPRL2024} in the USA, GANIL in France \cite{Ackermann2017PS}, IMP in China \cite{Zhang2012CPL,Yang2013NIMPSSB,Gan2022EPJA}, and various other leading laboratories worldwide \cite{Back2017EPJ,Hinde2017EPJ}. These efforts will not only facilitate the search for new elements with even higher $Z$, but also enable the production of previously synthesized isotopes at much higher rates, which is essential for more comprehensive studies of their structure. Complementing these experimental efforts, theoretical studies have advanced in parallel, aiming to unravel the underlying structure of SHN.

On the theoretical front, the study of SHN has predominantly utilized either macroscopic-microscopic models \cite{Brack1972RMP,Moller1994JPG,Moller1995ADNDT,Sobiczewski2007PPNP,Kowal2010PRC,Jachimowicz2010IJMPE,
Jachimowicz2011PRC,Wang2014PLB, Moller2015PRC,Moller2016ADNDT,Pomorski2018PRC,Chai2018CPC, Kostryukov2021CPC,Pomorski2022EPJA,Wang2024CPC} or nuclear density functional theory (DFT) \cite{Bender2003RMP,Meng2006PPNP,Adamian2021EPJA}. The advantages of using DFT include the intuitive interpretation of results in terms of single-particle states and intrinsic shapes, calculations performed in the full model space of occupied states, and the universality that enables their applications to all nuclei throughout the periodic chart. The latter feature is especially important for extrapolations to regions of short-lived nuclei, where few or even no data are available. Since the late 1990s, nuclear DFT has been systematically applied to investigate the structure of SHN \cite{Cwiok1996NPA,Lalazissis1996NPA,Bender1998PRC,Bender1999PRC,Cwiok1999PRL,Berger2001NPA,Goriely2002PRC,Burvenich2004PRC,Cwiok2005nature,Sheikh2009PRC,Erler2011JPG,Litvinova2011PRC,Litvinova2012PRC,Abusara2012PRC,
Warda2012PRC,Agbemava2015PRC,Baran215NPA,Agbemava2017PRC,Agbemava2019PRC,Olsen2019PRC,Taninah2020PRC,Akhilesh2021NPA,Jain2022NPA,Guo2024PRC,Jain2024PPNL,Alsultan2024NPA,Guo2024PRC,He2024PRC,Zhang2021PRC,Zhang2024YXPRC,
Reinhard2018EPJA,Zhou2016PS,Xu2024PLB,Giuliani2018PRC,Theeb2024NPA,Stone2019PRC,Meng2019SCP,Rodriguez2020EPJA,Sarriguren2021PLB,Nishu2022NPA,Wang2022CPC,Baran215NPA}.
Observables such as binding energies, deformations, $\alpha$-decay energies and half-lives, fission barriers and spontaneous-fission half-lives, fission isomers, and single-nucleon shell structure of SHN have been described successfully based on the non-relativistic Skyrme and Gogny density functionals as well as the relativistic density functionals.

In general, the static mean-field approximation is adopted in the studies based on DFT for SHN. The mean-field approximation is characterized by the breaking of symmetries in the underlying Hamiltonian, including translational, rotational, and particle number symmetries, which leads to the inclusion of static correlations such as deformations and pairing. To incorporate the dynamical correlations associated with quantum shape fluctuations (QSFs), one has to go beyond mean-field approximation by restoring the broken symmetries and mixing configurations of symmetry-breaking product states. In the last decade, DFT-based models that incorporate QSFs self-consistently have been employed to investigate ground state properties and low-lying spectra of SHN. A five-dimensional collective Hamiltonian (5DCH) \cite{Libert1999PRC,Prochniak2004NPA,Niksic2009PRC,Hinohara2010PRC} with parameters derived from DFT using the relativistic DD-PC1 functional \cite{Niksic2008PRC}, has been utilized to assess the effects of QSFs on the $Q_\alpha$ for the $\alpha$-decay chains $^{298}$120 and $^{300}$120 \cite{Prassa2012PRC}. Furthermore, this model is adopted to study the low-energy spectra and shape transitions within the isotopic chains spanning Fm to Fl \cite{Prassa2013PRC}. The isotopic dependence of various observables that characterize the transition between axially symmetric rotors and $\gamma$-soft rotors has also been investigated. Recently, we have analyzed how QSFs influences the ground state and fission properties of selected superheavy isotopic and isotonic chains, using both the PC-PK1 and DD-PC1 functionals \cite{Shi2019PRC}.

The generator coordinate method (GCM) based on the axially symmetric Hartree-Fock-Bogoliubov framework with the Skyrme SLy4 functional has been employed to calculate the dynamical correlation energies associated with QSFs for actinides and SHN with neutron numbers ranging from 136 to 184 \cite{Bender2013JPCS}. The influence of QSFs on the $Q_\alpha$ values is particularly significant near the $N = 184$ shell closure, where there is a pronounced variation in the dynamical correlation energies \cite{Bender2013JPCS}. Recent GCM calculations, which incorporate triaxial \cite{Egido2020PRL,Egido2021PRL} and octupole \cite{Guzman2021PRC} shapes based on Gogny energy density functionals, have been performed for selected superheavy isotopic chains and $\alpha$-decay chains. A novel type of shape coexistence between two distinct triaxial shapes has been predicted to occur in $^{290}$Fl \cite{Egido2020PRL}. For nuclei with $186 \leqslant N \leqslant202$, the coupling between quadrupole and octupole modes is found to be weak \cite{Guzman2021PRC}.

The aforementioned studies have demonstrated the significant impact of QSFs on the properties of SHN and the necessity of including a degree of freedom of the triaxial shape. However, these investigations mainly concentrate on a limited range of SHN. In this work, we aim to perform a systematic study on even-even SHN with $104 \leqslant  Z \leqslant 126$ and  $N\leqslant 258$ utilizing the five-dimensional collective Hamiltonian based on the triaxial relativistic DFT with the PC-PK1 functional \cite{Zhao2010PRC}. We mainly analyze the influence of QSFs on various properties, including binding energies, quadrupole deformations, the excitation energies of $2^+_1$, the energy ratios $R_{42}$, the transition probabilities $B(E2; 2^+_{1}\to 0^+_{1})$, two-nucleon separation energies, and $\alpha$-decay energies $Q_\alpha$. Notably, we report for the first time the absence of ground state $0^+$ in the transitional regions of SHN induced by the QSFs, despite the mean-field minima all exist in these nuclei. Moreover, QSFs smooth out the mean-field neutron shell closures at $N=184$ and $N=258$, shifting them to $N=182$ and $N=256$, respectively.

\section{The microscopic collective Hamiltonian model}\label{Theory}

Following our previous work \cite{Niksic2009PRC,Li2009PRC}, we adopt the microscopic DFT-based five-dimensional collective Hamiltonian (5DCH) to describe the QSFs related to triaxial shapes, namely the quadrupole vibrations, rotations, and the coupling of these collective modes. It takes the following form,
\begin{equation}\label{5DCH}
 \hat{H} (\beta, \gamma,\Omega) =\hat{T}_{\text{vib}}+\hat{T}_{\text{rot}} +V_{\text{coll}},
\end{equation}
with
 \begin{align}
\hat{T}_{\text{vib}}=&-\frac{\hbar^2}{2\sqrt{wr}} \left\{\frac{1}{\beta^4}  \left[\frac{\partial}{\partial\beta}\sqrt{\frac{r}{w}}\beta^4  B_{\gamma\gamma} \frac{\partial}{\partial\beta}\right.\right.\nonumber\\
& \left.\left.- \frac{\partial}{\partial\beta}\sqrt{\frac{r}{w}}\beta^3 B_{\beta\gamma}\frac{\partial}{\partial\gamma} \right]+\frac{1}{\beta\sin{3\gamma}} \left[    -\frac{\partial}{\partial\gamma} \right.\right.\\
& \left.\left.\sqrt{\frac{r}{w}}\sin{3\gamma}  B_{\beta \gamma}\frac{\partial}{\partial\beta} +\frac{1}{\beta}\frac{\partial}{\partial\gamma} \sqrt{\frac{r}{w}}\sin{3\gamma} B_{\beta \beta}\frac{\partial}{\partial\gamma} \right]\right\}\nonumber,
\end{align}
 \begin{align}
\hat{T}_{\text{\text{\text{rot}}}} =&\frac{1}{2}\sum_{k=1}^3{\frac{\hat{J}^2_k}{\mathcal{I}_k}}.
\end{align}
$\hat{J}_k$ denotes the components of the angular momentum in the body-fixed frame of a nucleus. Both the mass parameters $B_{\beta\beta}$, $B_{\beta\gamma}$, $B_{\gamma\gamma}$ and the moments of inertia $\mathcal I_k$ depend on the quadrupole deformation variables $\beta$ and $\gamma$. Two additional quantities  $r=B_1B_2B_3$ ($B_k$ calculated from $\mathcal I_k$ \cite{Niksic2009PRC}) and $w=B_{\beta\beta}B_{\gamma\gamma}-B_{\beta\gamma}^2 $, determine the volume element in the collective space. $V_{\text{coll}}$ is the collective potential that is calculated from the total mean-field energy subtracting the zero-point energy (ZPE) corrections $\Delta V_{\rm ZPE}$ \cite{Girod1979NPA}: $V_{\text{coll}}=E_{\rm mf}-\Delta V_{\rm ZPE}$. The full dynamics of 5DCH is determined by the seven deformation-dependent parameters, which are calculated using the perturbative cranking formula \cite{Girod1979NPA,Xiang2024PRC} based on the constrained triaxial relativistic Hartree-Bogoliubov (TRHB) model \cite{Niksic2010PRC}.

The corresponding eigenvalue equation is solved by expanding the eigenfunctions on a complete set of basis functions that depend on the deformation variables $\beta$ and $\gamma$, and the Euler angles \cite{Prochniak2004NPA}. And thus we obtain the energy spectrum $E^I_\alpha$ and collective  wave functions
\begin{align}
    \Psi^{IM}_\alpha(\beta, \gamma, \Omega)=\sum_{K\in\Delta I}\psi^I_{\alpha K}(\beta, \gamma)\Phi^I_{MK}(\Omega),
\end{align}
where $M$ and $K$ are the projections of angular momentum $I$ on the third axis in the laboratory and intrinsic frames, respectively, and $\alpha$ denotes the other quantum number. The probability density distribution of the collective state in the $(\beta, \gamma)$ plane, which takes the following form:
\begin{equation}
    \rho_{I \alpha}(\beta, \gamma)=\sum_{K \in \Delta I}\left|\psi_{\alpha K}^I(\beta, \gamma)\right|^2 \beta^3,
\end{equation}
could give a further insight into the shape fluctuation \cite{Majola2019PRC,Shi2018PRC}. The average values of the deformation parameters $\beta$ and $\gamma$ in the state $|I\alpha\rangle$ are calculated from
\begin{align}
\langle\beta\rangle_{I \alpha} & =\sqrt{\left\langle\beta^2\right\rangle_{I \alpha}} , \\
\langle\gamma\rangle_{I \alpha} & =\frac{1}{3} \arccos \frac{\left\langle\beta^3 \cos 3 \gamma\right\rangle_{I \alpha}}{\sqrt{\left\langle\beta^2\right\rangle_{I \alpha}\left\langle\beta^4\right\rangle_{I \alpha}}}.
\end{align}

\section{Results and discussion}\label{Results}

\begin{figure}[hbt]
    \centering
       \includegraphics[scale=0.4]{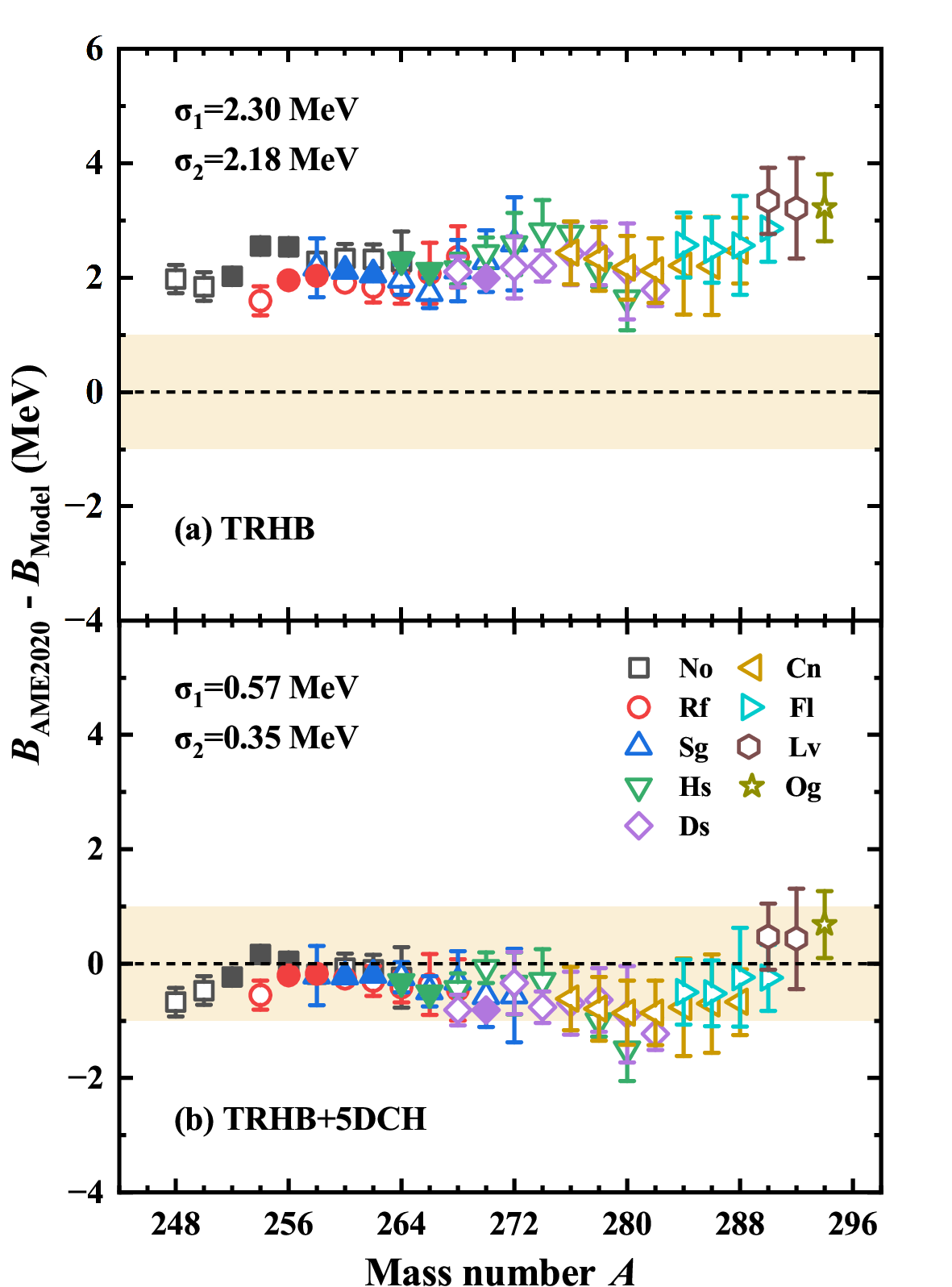}
    \caption{Binding energy difference between the AME2020 data \cite{Wang2021CPC} and (a) the TRHB calculations, and (b) the TRHB+5DCH for superheavy even-even nuclei with data available ($102 \leqslant Z \leqslant 118$).  $\sigma_1$ represents the root-mean-square deviation from the available 56 measured (solid symbols) and empirical (open symbols) masses, whereas $\sigma_2$ pertains solely to the 10 measured masses.}
    \label{Mass}
\end{figure}

\begin{figure}[bht]
    \centering{\includegraphics[width=0.5\textwidth]{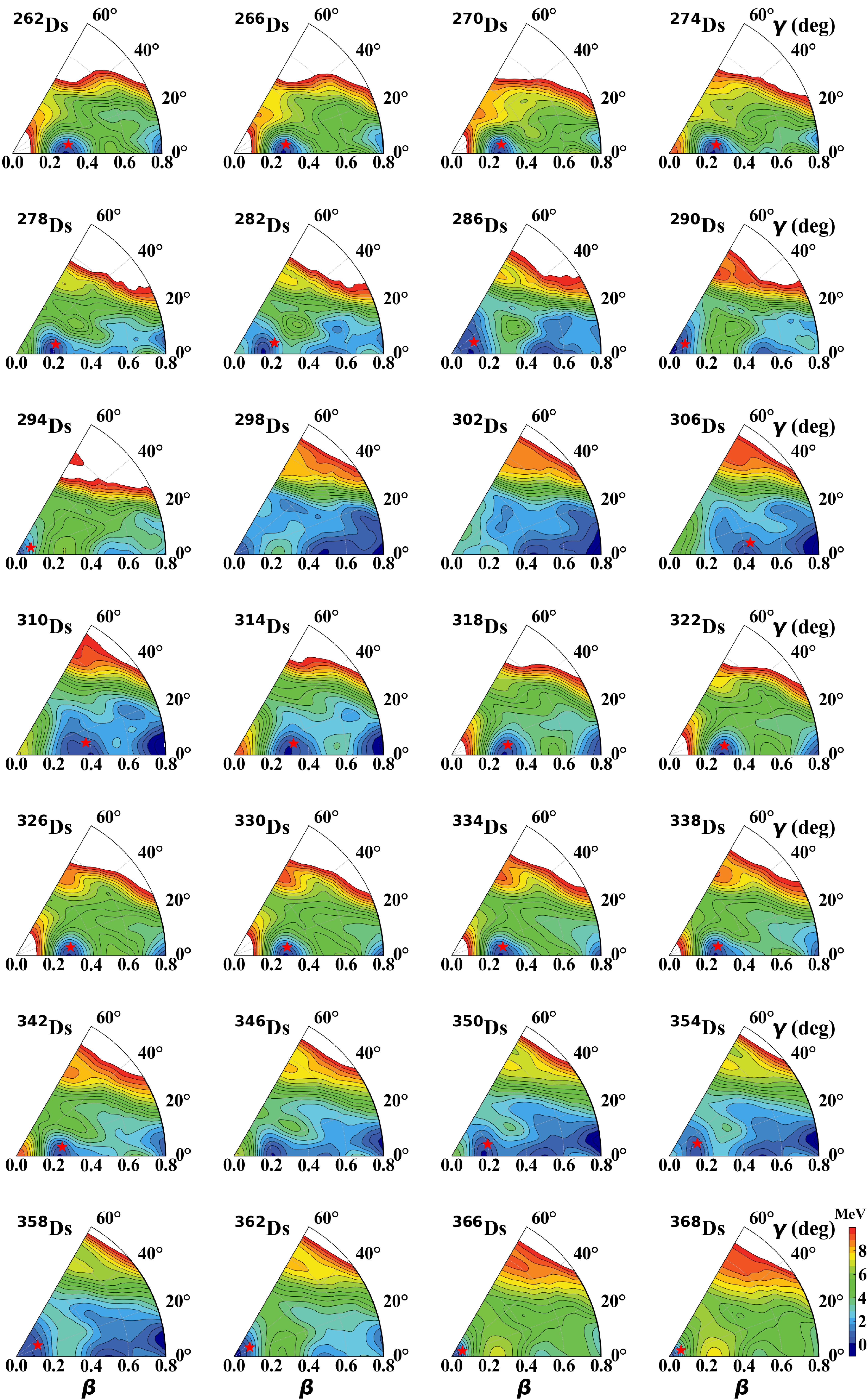}}
    \caption{\label{Etot-Ds}(Color online) Collective potentials in the $(\beta, \gamma)$ plane for the even-even Ds ($Z=110$) isotopic chain calculated by the constrained TRHB  with PC-PK1 functional. All energies are normalized with respect to the binding energy of the lowest minimum within the fission barrier. The energy difference between adjacent contour lines is 0.5 MeV. The red stars indicate the average deformations of the ground states $0^+_1$.}
\end{figure}

In this study, the PC-PK1 density functional \cite{Zhao2010PRC} is employed for the particle-hole channel in the TRHB calculation, combined with a finite-range separable pairing force \cite{Tian2009PLB} in the particle-particle channel. The TRHB equation is solved by expanding the quasiparticle wave functions in terms of a three-dimensional harmonic oscillator basis in Cartesian coordinates \cite{Niksic2010PRC}. For SHN with $104 \leqslant Z \leqslant 126$, the harmonic oscillator basis encompasses 18 major shells to ensure results are converged \cite{Shi2019PRC}. A large-scale constrained TRHB calculation is conducted to produce the collective potential and collective parameters in the $(\beta, \gamma)$ plane for the 5DCH. The cutoff for $\beta$ is set at $\beta_{\rm max}=0.8$ with a step size of $\Delta\beta=0.05$, while the range for $\gamma$ is from $0^{\rm o}$ to $60^{\rm o}$ with a step size of $\Delta\gamma=6^{\rm o}$. We have checked the numerical convergence of the results with a denser grid \( (\Delta\beta,\Delta\gamma) = (0.02, 2^\circ) \) and found that all calculated observables are well converged  \cite{SuppMat}.

In our previous work \cite{YangYL2021PRC}, we have employed the same model to systematically investigate the nuclear landscape with $Z\leqslant104$. Our calculations demonstrated significant improvements in predicting binding energies for even-even nuclei across the nuclear chart, particularly for medium and heavy mass regions, when compared with earlier studies using DD-PC1 and TMA functionals. The inclusion of QSFs played a crucial role in achieving remarkable agreement with experimental data, reducing the deviations to within $\pm1$ MeV for most nuclei. Moreover, the model can also provide good description of the inner fission saddle heights for the actinide nuclei \cite{Zhao2010PRC,Lu2014PRC,SuppMat}. Here we adopt the model to study the even-even SHN, mainly analyzing the influence of QSFs on the binding energies, quadrupole deformations, excitation energies and transition probabilities.

Firstly, we follow Ref. \cite{Zhang2021PRC} to evaluate the predictive power of the model in the superheavy region by comparing the theoretical binding energies with the available data \cite{Wang2021CPC, Kondev2021CPC, Huang2021CPC}. This evaluation is particularly crucial for extrapolating to the extremely heavy region ($Z=126$ and $N=258$). Figure \ref{Mass} illustrates the difference in binding energy between the theoretical results and the latest Atomic Mass Evaluation (AME2020) data \cite{Wang2021CPC} for superheavy even-even nuclei ($102 \leqslant Z \leqslant 118$). In the upper panel, the TRHB model shows a significant underestimation of nuclear binding energies, exceeding 2 MeV for most nuclei, with a root-mean-square deviation of 2.30 MeV. In contrast, the 5DCH results that account for the effect of QSFs, remain well within the range of $\pm 1$ MeV and the root-mean-square deviation is significantly reduced to 0.57 MeV, even lower at 0.35 MeV for the 10 measured masses. The 5DCH based on PC-PK1 functional effectively captures the isospin dependence of nuclear binding energies across isotopic chains and demonstrates consistent accuracy as $Z$ increases. This performance is significantly better than the well-calibrated macroscopic-microscopic mass models, such as WS4 ($\sigma_1=1.36$ MeV) \cite{Wang2014PLB} and FRDM ($\sigma_1=2.83$ MeV) \cite{Moller2012ADADT}, thus demonstrating the predictive power of the current model.  In the following, we will utlize the model to perform systematic calculations for SHN with $104 \leqslant  Z \leqslant 126$ and  $N\leqslant 258$.

To provide a clear understanding of shape evolution in this mass region, we present the collective potentials in the $(\beta, \gamma)$ plane calculated by constrained TRHB for the even-even isotopic chain of Ds ($Z=110$) in Fig. \ref{Etot-Ds}. Collective potentials for other isotopic chains are illustrated in the Supplemental Materials \cite{SuppMat}. $^{294,368}$Ds exhibit very deep spherical minima, corresponding to the spherical neutron shell closures at $N=184$ and $258$, respectively. While, isotopes with mass numbers $A=262\sim274$ and $314\sim342$, located in the middle shell, show significant prolate deformation with $\beta \approx 0.25-0.3$. For the transitional isotopes $^{278-290, 298-310, 346-362}$Ds, coexistence of two or more local minima are predicted, and one notes that the minima are soft with respect to either $\beta$ or $\gamma$. Notably, the shape evolution is consistent with those obtained from HFB calculations using the Gogny D1S functional \cite{Gogny,Delaroche2010PRC}. Starting from the collective potentials, we can construct a microscopic 5DCH that includes QSFs and allows for the calculation of the ground state $0^+_1$ denoted by the red stars in Fig. \ref{Etot-Ds}. The position of red star indicates the average deformations of the corresponding ground state, primarily following the well-developed minima. However, it is important to highlight that for some transitional nuclei $^{298,302,346}$Ds, ground states cannot be found within their minima, which tend to be quite shallow along the fission path.

\begin{figure}[htb]
    \centering
       \includegraphics[scale=0.3]{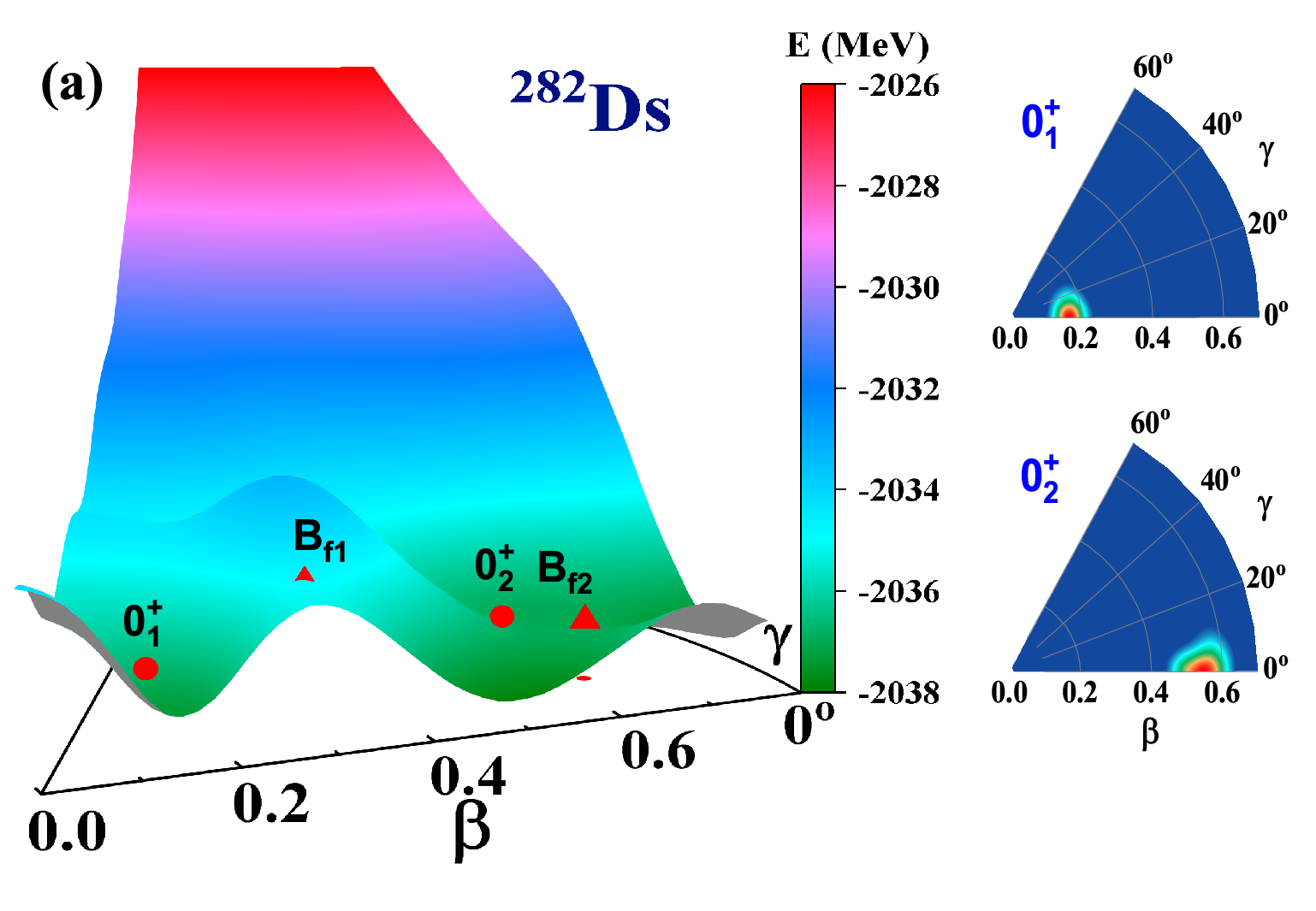}
       \includegraphics[scale=0.3]{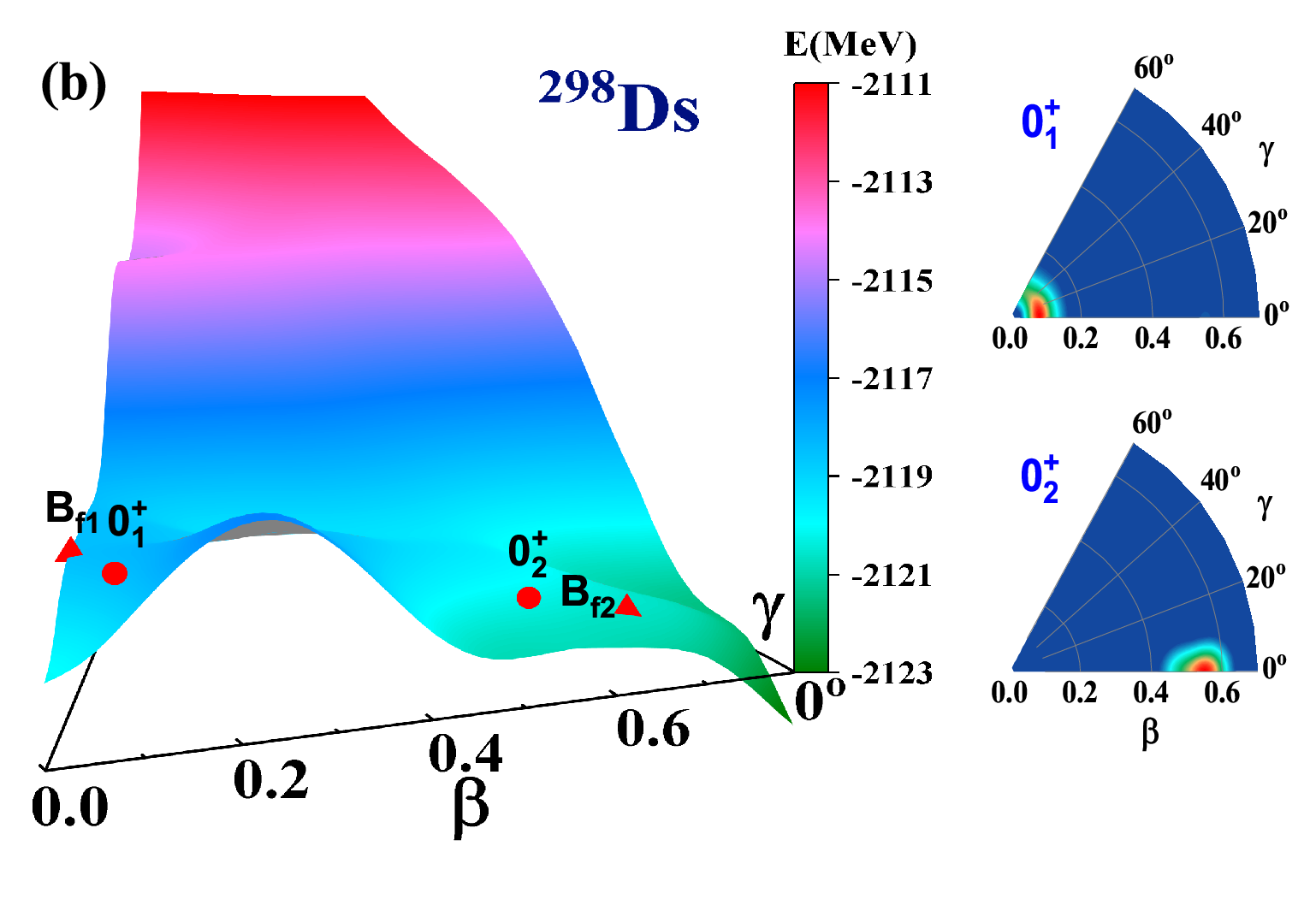}
    \caption{Three-dimensional plots of collective potential $V_{\rm coll}$ for $^{282}$Ds (a) and $^{298}$Ds (b) calculated using the constrained TRHB method with the PC-PK1 functional. Candidate $0^+$ states for each local minimum are marked by red solid dots, positioned at the average deformations of the corresponding state. The adjacent fission saddle is indicated by a red triangle. Probability density distributions in the ($\beta, \gamma$) plane for the candidate $0^+$ states are shown on the right of the panels, with red representing the peak density.
    \label{298Ds}}
\end{figure}

\begin{table}[htbp]
    \tabcolsep=6pt   
    \begin{center}
    \caption{Quadrupole deformations $(\beta, \gamma)$ and collective potential energies $V_{\rm coll}$ of the two minima, as well as the adjacent fission saddles $B_f$ for $^{282}$Ds and $^{298}$Ds shown in Fig. \ref{298Ds} calculated by the TRHB with PC-PK1 functional. Energies of candidate $0^+$ states $E(0^+)$ are calculated by 5DCH. All the energies are in the units of MeV.}
    \begin{tabular}{ccccc}
    \hline\hline
       &   \multicolumn{2}{c}{$^{282}$Ds} & \multicolumn{2}{c}{$^{298}$Ds} \\
       & Min1 & Min2 & Min1 & Min2 \\
    \hline
    $(\beta, \gamma) $   &  (0.17,$0^{\rm o}$)  &  (0.51, $0^{\rm o}$)     &   (0.00, $0^{\rm o}$)     &   (0.47, $0^{\rm o}$)  \\
    $V_{\rm coll}^{\rm min}$ & -2037.65 & -2038.15  &   -2119.93  &  -2120.92 \\
    $E(0^+)$   & -2036.42 & -2036.42  & -2118.26  &-2120.06  \\
    $B_f$      & -2034.94 & -2036.76 & -2118.42 &  -2120.80   \\
    $B_f-E(0^+)$ & 1.48 & -0.34 & -0.16 & -0.74  \\
    $B_f-V_{\rm coll}^{\rm min}$ & 2.71 & 1.39 & 1.51 & 0.12  \\
    $E(0^+)-V_{\rm coll}^{\rm min}$ & 1.23 & 1.73  & 1.67 & 0.86 \\

    \hline\hline
     \label{Ds-data}
    \end{tabular}
    \end{center}
    \end{table}

In Fig. \ref{298Ds}, we take $^{282}$Ds and $^{298}$Ds as examples to further illustrate the candidate $0^+$ states and fission saddles in their corresponding collective potentials. The probability density distributions in the $(\beta, \gamma)$ plane for these $0^+$ candidates are also shown. The detailed values of the deformations and energies for the minima, fission saddles, and $0^+$ candidates are summarized in Table \ref{Ds-data}. To more accurately identify the potential minima $V_{\rm coll}^{\rm min}$ and fission saddles $B_f$, we employ the immersion method \cite{PhysRevC.79.064304} on the finely interpolated collective potential $V_{\rm coll}$ which is constructed using the standard cubic spline technique with a very dense grid. The ground state $0^+_1$ of $^{282}$Ds locates at the minimum with $(\beta, \gamma) = (0.17, 0^{\rm o})$, exhibiting an energy of $-2036.42$ MeV, which is lower than the adjacent fission saddle with $B_{f1} = -2034.94$ MeV. However, in the local minimum at $(\beta, \gamma) \approx (0.51, 0^{\rm o})$, no $0^+$ states are found to be lower than the adjacent fission saddle ($B_{f2}$) located at $(\beta, \gamma) \approx (0.62,5^{\rm o})$. The 5DCH calculations predict only a quasi-localized state with $E(0^+) = -2036.42$ MeV, and it has been confirmed that both the energy of this state and its probability density distribution remain stable when the deformation cutoff $\beta_{\rm max}$ is increased in the 5DCH calculation. This may manifest itself as a resonance  \cite{hazi1970stabilization,Zhang2007RealSM} during the synthesis of this isotope, but it will undergo fission on a timescale similar to that of induced fission \cite{Bender2020JPGNPP,Schunck2016RPP,SuppMat}. Similarly, in $^{298}$Ds, both candidates for the $0^+$ states are located above the adjacent fission saddles, indicating that no ``bound'' $0^+$ states can be identified for either local minimum. This implies that this nucleus may not be realizable with current experimental techniques.

From the last two rows in Table \ref{Ds-data}, we note that whether the $0^+$ states are ``bound'' or not is determined by the competition between the height of the fission saddle and the zero-point vibrational energy defined as $E(0^+)-V_{\rm coll}^{\rm min}$ in 5DCH. The zero-point vibrational energy is induced by the quantum fluctuation in the collective shape space, which is mainly related with the stiffness of the collective potential \cite{Shi2019PRC}. Both nuclei exhibit small zero-point vibrational energies and low fission saddles due to their transitional nature and soft collective potentials.  However, the relatively lower fission saddles lead to the ``unbound'' $0^+_2$ in $^{282}$Ds and $0^+_{1,2}$ in $^{298}$Ds.

\begin{figure}[ht]
\centering{\includegraphics[width=0.5\textwidth]{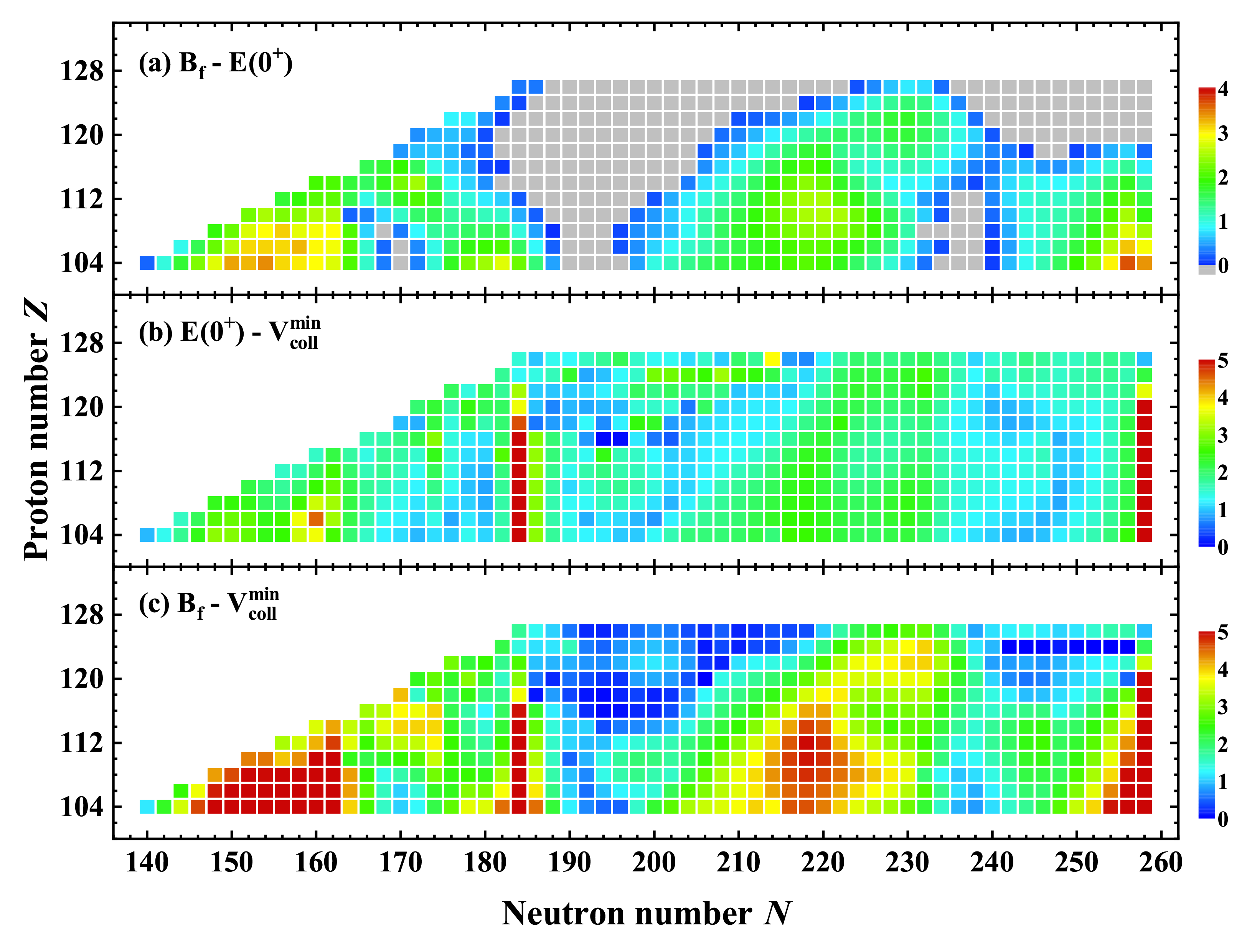}}
\caption{\label{SM} (Color online) (a) Energy difference between the fission saddle and the candidate $0^+$ state in the deepest minimum for each superheavy nucleus. The zero-point vibrational energies of the $0^+$ candidates and the heights of adjacent fission saddles are also shown in panels (b) and (c), respectively. All energies are in the units of MeV.}
\end{figure}

Furthermore, following the same approach as for the Ds isotopic chain, we systematically analyze 585 SHN with $104 \leqslant  Z \leqslant 126$ and $N\leqslant258$. The height of fission saddle and zero-point vibrational energy of the candidate $0^+$ state in the deepest minimum for each nucleus is shown in Fig. \ref{SM} (b) and (c), respectively, while in panel (a), we plot the difference between them. Our results identify two distinct ``unbound'' regions with $N\gtrsim 184$ and $N\gtrsim 240$. In addition, some lighter $Z$ nuclei with $N\sim170$ and $N\sim236$ are also predicted to be ``unbound''. The total count of such nuclei is 175, representing a significant proportion compared to the total of 585 even-even SHN calculated. While the zero-point vibrational energies of most ``unbound'' SHN remain modest ($0.3\sim2$ MeV), these nuclei exhibit significantly lowered fission saddles, particularly for nuclei with $Z\geqslant116$ where fission saddles nearly vanish. Such features of the fission saddles are consistent with those calculated from the non-relativistic Skyrme SV-min density functional \cite{Reinhard2018EPJA}, several commonly used relativistic density functionals \cite{Taninah2020PRC,SuppMat}, and macroscopic-microscopic models \cite{Moller2015PRC,Baran2005PRC,Kowal2010PRC,JachimowiczPRC2017}. Notably, the fission saddles for Rf ($Z=104$), Sg ($Z=106$), and Hs ($Z=108$) isotopes with $N\sim170$ are also predicted to be very low, $<2$ MeV. The competitive zero-point vibrational energies render these nuclei weakly bound or even ``unbound''. This is supported by the experiments that point to very short spontaneous fission half lives in this region \cite{Oganessian2015RPP}. Fortunately, the highly anticipated new elements $Z=119$ and $120$ ($N\approx178-180$), potentially synthesizable via fusion reactions, remain bound when accounting for QSFs. In the future, it might be feasible to synthesize neutron-rich SHN with $N\gtrsim184$, through other methods, such as multi-nucleon transfer reactions. In such scenarios, a careful assessment of QSF effects will be crucial.

\begin{figure*}[ht]
\centering{\includegraphics[width=0.72\textwidth]{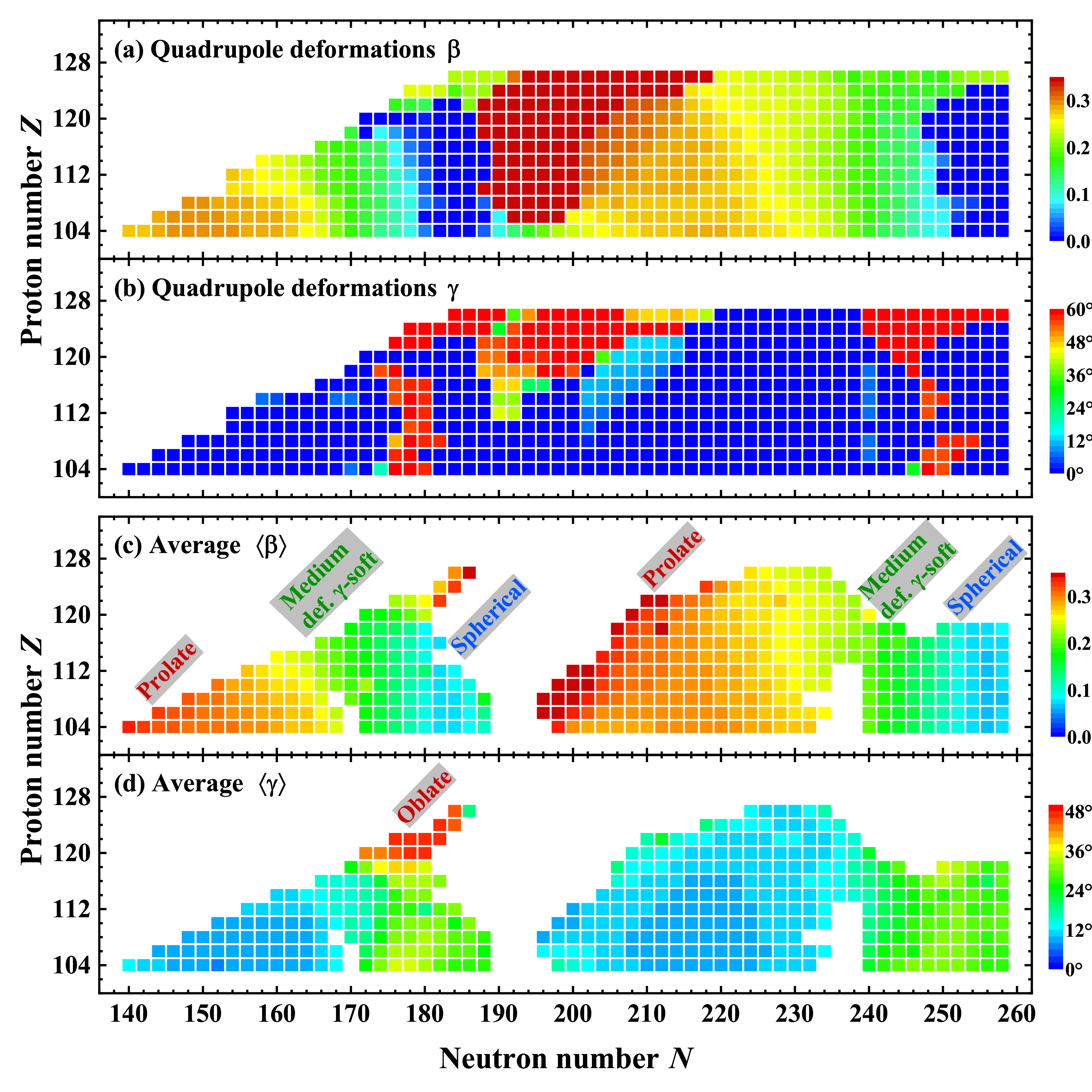}}
\caption{\label{Qd}(Color online) Quadrupole deformations $\beta$ (panel a) and $\gamma$ (panel b) of the mean-field minima calculated from TRHB, and average quadrupole deformations $\langle\beta\rangle$ (panel b) and $\langle\gamma\rangle$ (panel c) of ground states $0^+_1$ from 5DCH for the even-even SHN with  $104 \leq Z \leq 126$ from proton drip-line to $N=258$.}
\end{figure*}

In the following, we will analyze the impact of QSFs on the quadrupole deformations. Figure \ref{Qd} illustrates the quadrupole deformations $\beta$ and $\gamma$ of the mean-field minima obtained from TRHB, along with the average quadrupole deformations $\langle\beta\rangle$ and $\langle\gamma\rangle$ for the ground states $0^+_1$ obtained from 5DCH. The lower two panels focus on the SHN with ground states bounded by the fission saddles. Similar as the findings in Fig. \ref{Etot-Ds}, the mean-field calculations indicate significant prolate deformations for lighter elements ($Z=104\sim 118$) with neutron numbers $N=140\sim160$ and $N=188\sim 220$. This is accompanied by a shape transition to medium deformations and then weakly oblate shapes at $N\approx178$ and $250$, finally approaching to spherical nuclei around $N=184$ and $N=258$. Medium to large oblate deformations are preferred for isotopes with $Z\geqslant 120$. The shape evolution of SHN is generally consistent with that from other DFT calculations  \cite{Prassa2012PRC,Heenen2015NPA,Gogny,Agbemava2015PRC,Taninah2020PRC,Jain2022NPA,Guo2024PRC} and macroscopic-microscopic models \cite{Jachimowicz2011PRC,Heenen2015NPA,Moller2016ADNDT}. The discrepancies mainly lie in the predictions concerning the spherical neutron shell closure at $N=184$ or $N\approx 174$, as well as the region for oblate deformations \cite{Heenen2015NPA}. Furthermore, the emergence of large deformations ($\beta>0.35$) in SHN with $N\approx 190-210$ appears to be a distinguishing characteristic of numerous relativistic DFT calculations \cite{Agbemava2015PRC,Zhang2022ADNDT}. However, when accounting for QSFs, no truly bound ground states exist for these nuclei. For the bound nuclei in Fig. \ref{Qd} (c, d), the inclusion of QSFs leads to a more gradual shape evolution. In particular, the prolate-oblate shape phase transitions predicted in lighter $Z$ SHN ($N \approx 178$ and $250$) are smoothed out in the beyond-mean-field 5DCH calculations, whereas the oblate shape persists in higher $Z$ SHN ($N \approx 178$).


\begin{figure}[ht]
\centering{\includegraphics[width=0.5\textwidth]{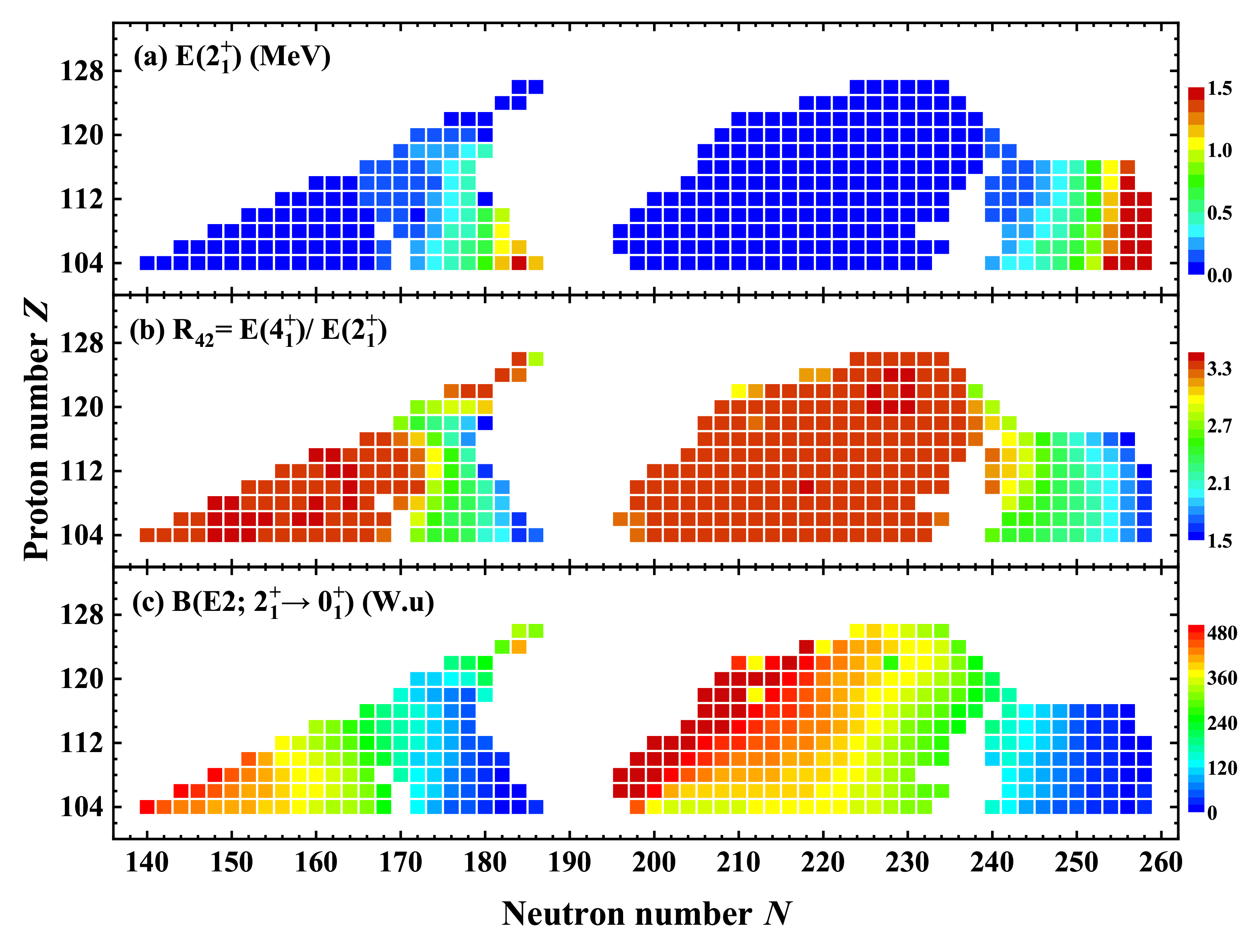}}
\caption{\label{R42-tem}(Color online) Excitation energies of $2^+_1$ states $E(2^+_1)$ (a), energy ratios $R_{42}=E(4^+_1)/E(2^+_1)$ (b), and $B(E2; 2^+_1\to 0^+_1)$ (c) for the even-even SHN with  $104 \leq Z \leq 126$ and $N\leqslant 258$ calculated by the 5DCH based on TRHB with PC-PK1 functional.}
\end{figure}


The shape transition is clearly illustrated in the variations of characteristic collective observables, such as the excitation energies of $2^+_1$ states $E(2^+_1)$, the energy ratios $R_{42}=E(4^+_1)/E(2^+_1)$, and $B(E2; 2^+_1\to 0^+_1)$, shown in Fig. \ref{R42-tem}. The excitation energies $E(2^+_1)$ for the spherical SHN around $N=184$ and 258 shells are rather high, $\gtrsim 1$ MeV, but decrease rapidly to $<100$ keV in the well deformed regions. We note that the $E(2^+_1)$ for $^{254,256}$Rf have been measured and the values are 48 and 44 keV, respectively, which are overestimated by the present model, 66 and 68 keV, respectively. This is because the moments of inertia in our present model are calculated by the Inglis-Beliaev formula, which do not include the Thouless-Valatin dynamical rearrangement contributions and, thus, would systematically underestimate the empirical values by $30\%\sim50\%$. However, we note that the calculated energy ratios $R_{42}$ are 3.35 and 3.34 for $^{254,256}$Rf, respectively, which are in good agreement with the experimental values: 3.29 and 3.36. Combined with our previous systematic calculations \cite{Quan2017PRC,Yang2023PRC}, these results confirm our model's ability to describe collective properties, particularly the energy ratios $R_{42}$ and $B(E2; 2^+_1\to 0^+_1)$. The systematic calculations in Fig. \ref{R42-tem} (b) and (c) indicate that the SHN with $N\lesssim 170$ and $196\lesssim N \lesssim 240$ are predicted to be close to a rotor characterized by $R_{42}\gtrsim 3.2$ and $B(E2; 2^+_1\to 0^+_1)\gtrsim $ 240 W.u., while the lighter $Z$ SHN with $N\sim 184$ and 258 are spherical characterized by $R_{42}\lesssim 2.0$ and low $B(E2; 2^+_1\to 0^+_1)$. In between, the SHN are transitional nuclei with $R_{42}\in (2.0, 3.0)$. The predicted collective properties are generally consistent with other theoretical calculations given by the Skyrme \cite{Giuliani2019RMP,Baran2013PS} and Gogny \cite{Delaroche2010PRC,Gogny} functionals.

\begin{figure}[ht]
    \centering{\includegraphics[width=0.45\textwidth]{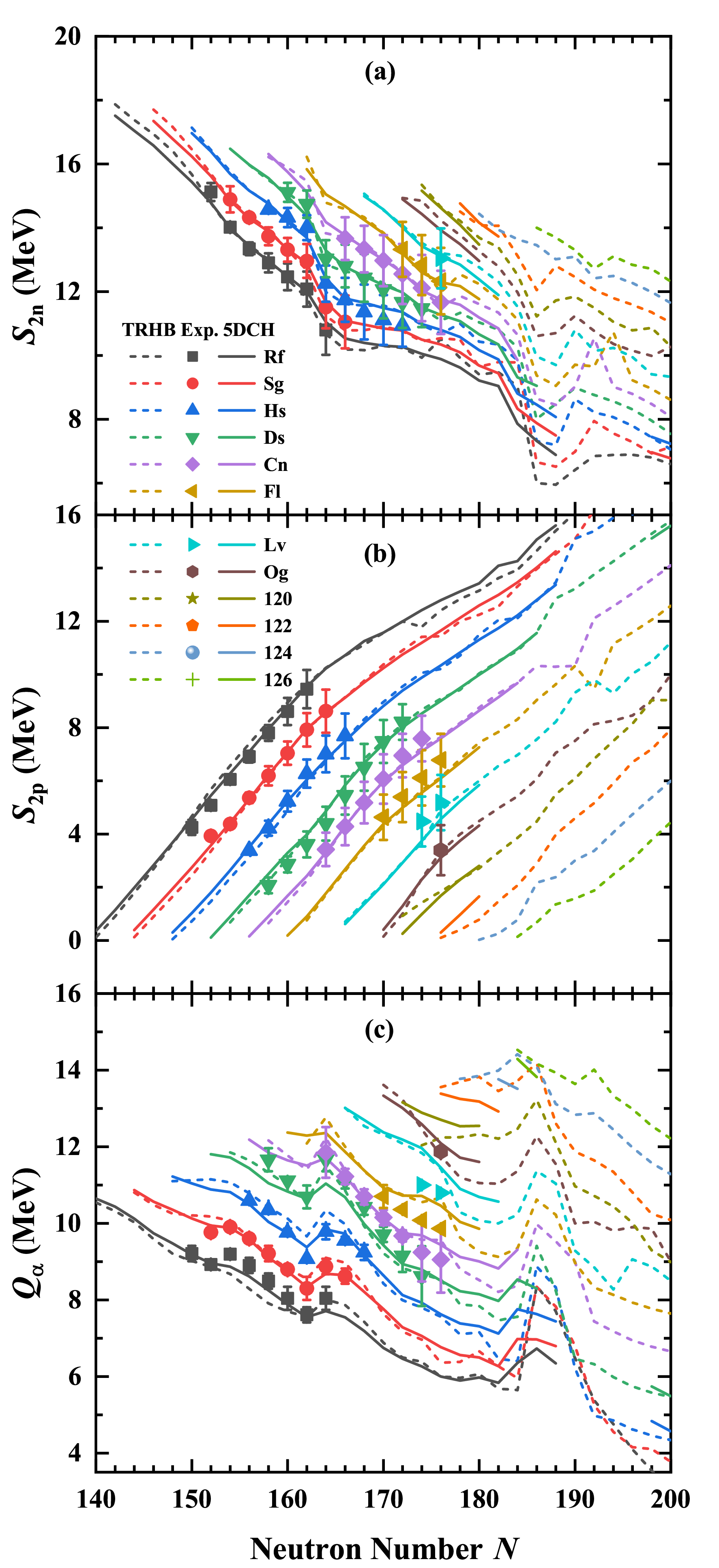}}
    \caption{\label{S2n}(Color online)  Two-neutron separation energies $S_{2n}$ (upper panels), two-proton separation energies $S_{2p}$ (middle panels), and $\alpha$ decay energies $Q_\alpha$ (lower panels) of SHN as functions of the neutron number calculated by the mean-field TRHB (dashed lines) and 5DCH (solid lines), in comparison with available data (solid symbols) \cite{Wang2021CPC}.}
\end{figure}

\begin{figure}[htb]
    \centering
       \includegraphics[scale=0.38]{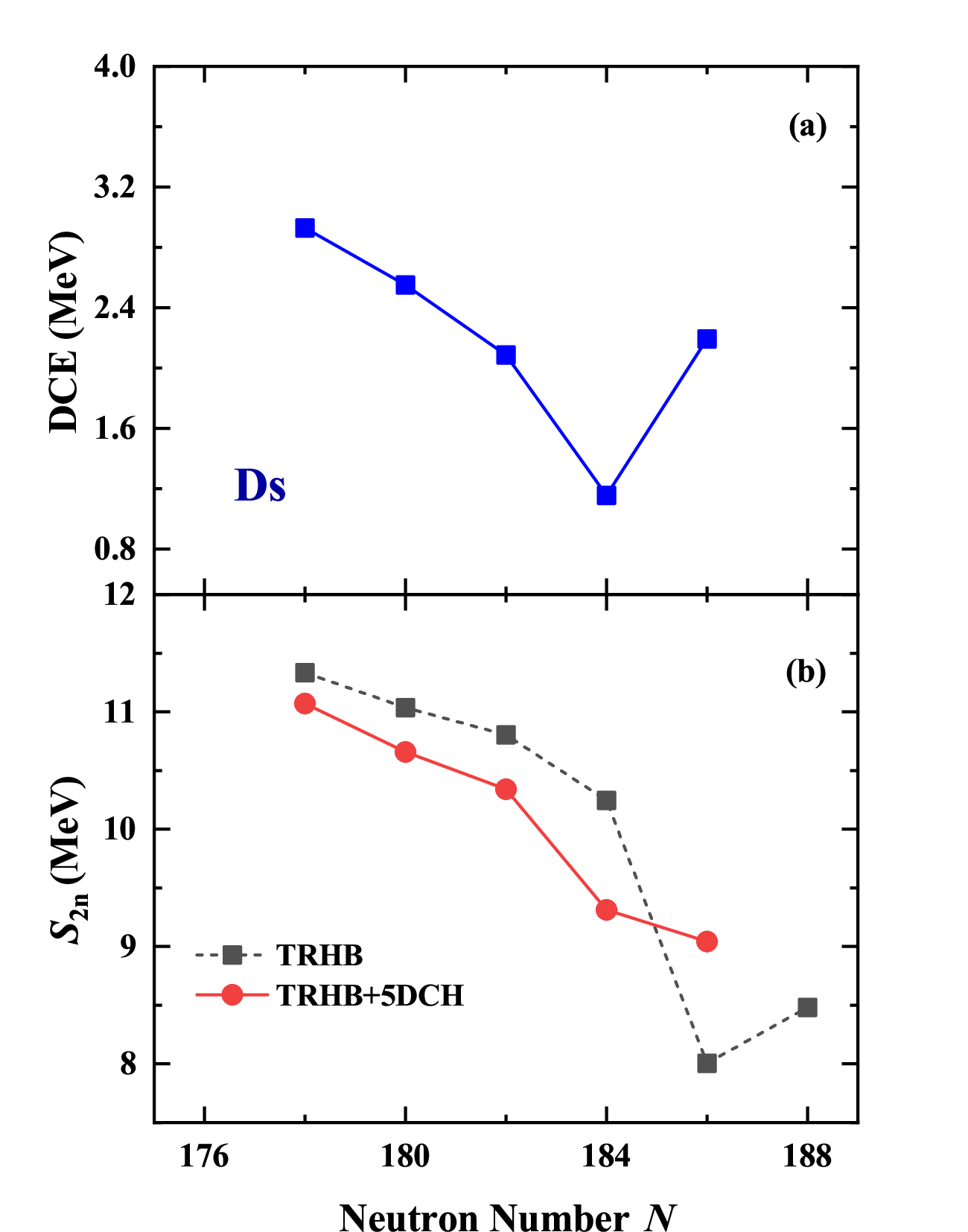}
    \caption{Dynamical correlation energies (upper) and two-neutron separation energies (lower) for the Ds ($Z=110$) isotopes near $N=184$ calculated by the 5DCH based on the TRHB calculations with PC-PK1 functional.}
    \label{Ds-shell}
\end{figure}

In Fig. \ref{S2n}, we examine the influence of QSFs on the two-neutron separation energies $S_{2n}$, two-proton separation energies $S_{2p}$, and $\alpha$-decay energies $Q_\alpha$ by comparing the mean-field TRHB and 5DCH results. Here we only show the results for SHN with $N\leqslant200$, and those for heavier isotopes can be found in the Supplemental Materials \cite{SuppMat}. The mean-field results agree well with the available data, yielding root-mean-square deviations of 0.19, 0.32, and 0.37 MeV for $S_{2n}$, $S_{2p}$, and $Q_\alpha$, respectively. Notably, sharp variations in $S_{2n}$ and $Q_\alpha$ occur when the neutron number crosses the deformed shell at $N=162$, consistent with other DFT calculations \cite{Agbemava2015PRC,Taninah2020PRC, Prassa2012PRC,Heenen2015NPA,Gogny, Agbemava2015PRC, Zhang2022ADNDT} and  macroscopic-microscopic models \cite{Heenen2015NPA}. Additionally, the discontinuities in both quantities become more pronounced near the neutron spherical shells at $N=184$ and $258$ \cite{SuppMat}. While $S_{2p}$ increases steadily by adding more neutrons. 

By incorporating QSFs, the 5DCH calculations smooth the evolution of quadrupole deformations (cf. Fig. \ref{Qd}), leading to more gradual variations in $S_{2n}$ and $Q_{\alpha}$ for ``bound'' SHN. This improves the description of the available data, with root-mean-square deviations of 0.17, 0.30, and 0.32 MeV, respectively. Interestingly, the sharp variations of $S_{2n}$ and $Q_\alpha$ at $N=184$ and $258$ in mean-field calculations are significantly reduced and shifted to $N=182$ and $256$ when QSFs are included \cite{SuppMat}. This phenomenon can be understood by analyzing the evolution of dynamical correlation energies associated with QSFs, as demonstrated in Fig. \ref{Ds-shell} for the Ds isotopes (Those for all the calculated SHN can be found in the Supplemental Materials \cite{SuppMat}). The dynamical correlation energy is defined as the energy difference between the mean-field minimum in the intrinsic frame and the ground state $0^+_1$ in the laboratory frame. In the medium and heavy mass regions \cite{Lu2015PRC,YangYL2021PRC,Delaroche2010PRC}, the calculated dynamical correlation energies exhibit a dip at the spherical shells, as the spherical nucleus only includes vibrational motion, while both vibrational and rotational motions emerge in the deformed nucleus. Overall, incorporating dynamical correlation energies mitigates the overestimated drops of $S_{2n}$ near spherical shells, resulting in a better agreement with the experimental data \cite{Sun2022CPC}. Similar evolution of dynamical correlation energies is found in Ds isotopes, a dip at the spherical shell $N=184$. However, in the SHN region, the less pronounced drop of $S_{2n}$ at $N=184$ in the mean-field calculation becomes more gradual when taking into account the dynamical correlation energies, and a new sharp drop emerges between $N=182$ and $N=184$. A similar mechanism accounts for the weakening of the $N=258$ shell.

\begin{figure}[ht]
    \centering{\includegraphics[width=0.5\textwidth]{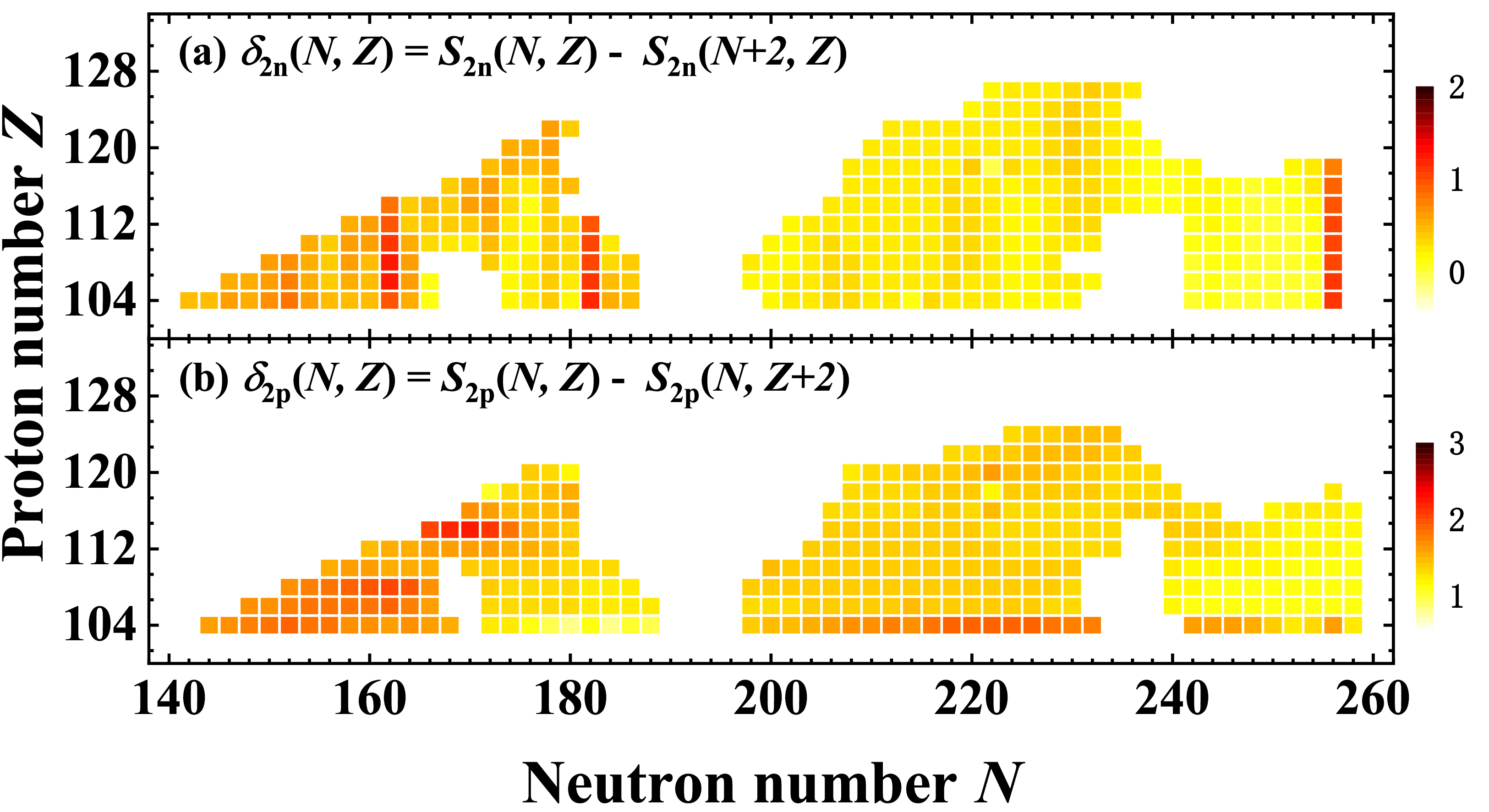}}
    \caption{\label{dS2n}(Color online) (a) Two-neutron gaps $\delta_{2n}$ and (b) two-proton gaps $\delta_{2p}$  of ``bound'' even-even nuclei with $104 \leqslant Z \leqslant 126$ calculated from 5DCH based on TRHB with PC-PK1 functional.}
\end{figure}

Finally, in Fig. \ref{dS2n}, we analyze the evolution of two-neutron gap $\delta_{2n}=S_{2n}(N, Z)-S_{2n}(N+2, Z)$ and the two-proton gap $\delta_{2p}=S_{2p}(N, Z)-S_{2p}(N, Z + 2)$ to identify the possible shell closures when considering the QSFs. Peaks are found at $N=162, 182$, and $256$ for $\delta_{2n}$, which are consistent with the findings in the $S_{2n}$ in Fig. \ref{S2n}. For the protons, $Z=114$ is a shell closure in the neutron-deficient SHN region, while the $Z=104$ shell emerges in the neutron-rich region.

\section{\label{Summary} Summary and outlook}

In conclusion, the even-even superheavy nuclei with $104 \leqslant  Z \leqslant 126$ and $N\leqslant258$ have been investigated using the microscopic 5DCH framework based on constrained TRHB calculations with the PC-PK1 relativistic density functional. The 5DCH calculations reproduce the  binding energies across the isotopic chains and show consistent accuracy as $Z$ increases, underscoring the predictive power of this model.

The mean-field calculations indicate significant prolate deformations for lighter elements ($Z=104\sim 118$) with neutron numbers $N=140\sim160$ and $N=188\sim 220$. This is accompanied by a shape transition to medium deformations and then weakly oblate shapes at $N\approx178$ and $250$, finally approaching to spherical nuclei around $N=184$ and $N=258$. Medium to large oblate deformations are preferred for isotopes with $Z\geqslant 120$. While, the 5DCH calculations, which take into account quantum shape fluctuations (QSFs), exhibit a smoother shape evolution. Remarkably, there are no $0^+$ ground state bounded by fission saddles in the transitional regions with $N\gtrsim184$ and $N\gtrsim240$, where both the spherical and deformed potential wells are all rather shallow and thus the corresponding fission saddles are relatively lower compared to the zero-point vibrational energies. A total of 175 such nuclei have been identified, accounting for a substantial proportion of the 585 even-even superheavy nuclei calculated. Fortunately, the highly anticipated new elements $Z=119$ and $120$ ($N\approx178-180$) that could be synthesized by fusion reactions are still ``bound'' when considering QSFs.

The shape transition is further investigated by examining the evolution of characteristic collective observables: $E(2^+_1)$, $R_{4/2}=E(4^+_1)/E(2^+_1)$, and $B(E2; 2^+_1\to 0^+_1)$. SHN with $N\lesssim 170$ and $196\lesssim N \lesssim 240$ are predicted to be close to a rotor characterized by $R_{42}\gtrsim 3.2$ and $B(E2; 2^+_1\to 0^+_1)\gtrsim $ 240 W.u., while the lighter $Z$ SHN with $N\sim 184$ and 258 are spherical characterized by $R_{42}\lesssim 2.0$ and low $B(E2; 2^+_1\to 0^+_1)$.

The 5DCH results are in good agreement with the available data for the $S_{2n}$, $S_{2p}$, and $Q_\alpha$, with the root-mean-square deviations of 0.17, 0.30, and 0.32 MeV, respectively. The sharp variations at $N=184$ and $258$ for $S_{2n}$ and $Q_\alpha$ predicted in the mean-field calculations are significlantly reduced and shifted to $N=182$ and $256$ in the 5DCH calculations, which is caused by the rapid evolution of the dynamical correlation energies related to QSFs around the nuclear shells.

It is important to highlight that the prediction for  ``unbound" SHN is highly sensitive to the height of the fission saddle, which is influenced by various factors, such as the treatment of pairing correlations and pairing strength, as noted in previous studies \cite{Tao2017PRC}. High-order deformations also play a role in this context. Notably, octupole deformations significantly contribute to the stability of superheavy nuclei by lowering their ground state minima \cite{Yang2022PRC,Xu2024PLB,Zhao2024PRC}. However, this effect mainly works in some specific regions, as referenced in the literature \cite{Agbemava2017PRC} and our calculations presented in the Supplemental Materials \cite{SuppMat}. Whereas, the ``unbound'' states is predicted in a wide region, particularly for larger $Z$ isotopes. Additionally, it is worth mentioning that octupole degrees of freedom generally decrease the height of fission barriers \cite{Sobiczewski2010IJMPE,Lu2012PRC,Lu2014PRC}. To determine the exact region for ``unbound''  SHN, further research is needed to incorporate at least octupole degrees of freedom into beyond-mean-field calculations. Recently, we have developed a seven-dimensional collective Hamiltonian (7DCH) that incorporates both triaxial quadrupole and octupole degrees of freedom \cite{xiang2025microscopic}, based on the multi-dimensional constrained relativistic DFT \cite{Lu2014PRC}. Consequently, we will be able to investigate this phenomenon further using the microscopic 7DCH.

\begin{acknowledgements}
We acknowledge helpful discussions with Prof. Jianmin Dong. This work was partly supported by the National Natural Science Foundation of China (Grants No. 12375126), the Fundamental Research Funds for the Central Universities. The PHD Foundation of Chongqing Normal University (No. 23XLB010), the Science and Technology Research Program of Chongqing Municipal Education Commission (No. KJQN202300509), Joint Fund Project of Education Department in Guizhou Province (No. Qian Jiao He KY Zi[2022]098).
\end{acknowledgements}

\bibliography{apssamp}

\vspace{5cm}
\begin{table*}
{\fontsize{20}{30} \selectfont \textbf{Supplemental Materials}}
\end{table*}
\begin{table*}[b]
    \tabcolsep=9pt   
    \begin{center}
    \renewcommand{\tablename}{TABLE}
    \caption{Numerical check for the grid used in the TRHB and 5DCH models. The resulting quadrupole deformations ($\beta,\  \gamma$) and energies ($V^{\rm min}_{\rm coll}$) of the collective potential minima, fission saddle energies ($B_f$), \(0^+\) energies ($E(0^+)$) from the 5DCH, and energy differences for $^{282}$Ds are summarized in the table.}
    \begin{tabular}{ccccc}
    \hline\hline
    & \multicolumn{2}{c}{Min1} & \multicolumn{2}{c}{Min2}  \\
    \hline
     &  $(\Delta\beta, \Delta\gamma)=(0.05, 6^{\rm o})$ &   $(\Delta\beta, \Delta\gamma)=(0.02, 2^{\rm o})$   &$(\Delta\beta, \Delta\gamma)=(0.05, 6^{\rm o})$  & $(\Delta\beta, \Delta\gamma)=(0.02, 2^{\rm o})$   \\
     \hline
     ($\beta,\gamma$) & $(0.17, 0^{\rm o})$ &  $(0.17, 0^{\rm o})$  &  $(0.51, 0^{\rm o})$  &  $(0.51, 0^{\rm o})$   \\
     $V_{\rm coll}^{\rm min}$  & -2037.65  &  -2037.66  &  -2038.15  &  -2038.15   \\
     $E(0^+)$  & -2036.42 &  -2036.42  & -2036.42 & -2036.40 \\
     $B_f$  &  -2034.94 &  -2034.97  & -2036.76 & -2036.74\\
     $B_f - E(0^+)$  & 1.48 & 1.45 & -0.34 & -0.34\\
     $B_f - V_{\rm coll}^{\rm min}$ & 2.71  &  2.69  &  1.39  &  1.41   \\
      $E(0^+) - V_{\rm coll}^{\rm min}$  &1.23 & 1.24  & 1.73 &1.75\\
    \hline\hline
     \label{Tables4}
    \end{tabular}
    \end{center}
    \end{table*}

\begin{table*}[b]
    \tabcolsep=9pt   
    \begin{center}
    \renewcommand{\tablename}{TABLE}
    \caption{Same as TABLE. \ref{Tables4}, but for $^{298}$Ds.}
    \begin{tabular}{ccccc}
    \hline\hline
    & \multicolumn{2}{c}{Min1} & \multicolumn{2}{c}{Min2}  \\
    \hline
    &  $(\Delta\beta, \Delta\gamma)=(0.05, 6^{\rm o})$ &   $(\Delta\beta, \Delta\gamma)=(0.02, 2^{\rm o})$   &$(\Delta\beta, \Delta\gamma)=(0.05, 6^{\rm o})$  & $(\Delta\beta, \Delta\gamma)=(0.02, 2^{\rm o})$   \\
    \hline
     ($\beta,\gamma$) & $(0.00, 0^{\rm o})$  &  $(0.00, 0^{\rm o})$  &  $(0.47, 0^{\rm o})$  &  $(0.47, 0^{\rm o})$   \\
     $V_{\rm coll}^{\rm min}$  &  -2119.93 &  -2119.93  &  -2120.92  &  -2120.93   \\
    E($0^+$) & -2118.26 &  -2118.26  & -2120.06 & -2120.03\\
     $B_f$  &  -2118.42 &  -2118.40  & -2120.80 & -2120.79 \\
    $B_f - E(0^+)$  & -0.16 & -0.14  & -0.74 & -0.76 \\
     $B_f - V_{\rm coll}^{\rm min}$ & 1.51  &  1.53  &  0.12  &  0.14   \\
      $E(0^+) - V_{\rm coll}^{\rm min}$ & 1.67 & 1.67 & 0.86 & 0.90 \\
    \hline\hline
     \label{Tables5}
    \end{tabular}
    \end{center}
    \end{table*}

\begin{table*}[b]
\tabcolsep=14pt   
\begin{center}
\caption{Energies of fission saddles $E_f$ and candidates of $0^+$ states $E(0^+)$, as well as their difference for $^{298-308}$Ds calculated by the five-dimensional collective Hamiltonian  based on the constrained triaxial relativistic Hartree-Bogoliubov model with the PC-PK1 \cite{Zhao2010PRC}, PCF-PK1 \cite{Zhao2022PRC}, and DD-PC1 \cite{Niksic2008PRC} functionals. All the energies are in the unit of MeV.}
\begin{tabular}{ cccccccc}
\hline\hline
& & $^{298}$Ds & $^{300}$ Ds& $^{302}$Ds & $^{304}$Ds & $^{306} $Ds & $^{308}$Ds \\
\hline
        & $B_f$ & -2118.42 & -2129.54 & -2138.08 & -2146.23 & -2154.08 & -2161.72 \\
PC-PK1  & $E(0^+)$ & -2118.26 & -2128.95 & -2137.54 & -2145.9 & -2154.05 & -2161.98 \\
        & $B_f-E(0^+)$ & -0.16 & -0.59 & -0.54& -0.33 & -0.03 & 0.26
\vspace{0.2cm}\\
        & $B_f$ & -2113.93 & -2124.37 & -2132.5 & -2140.31 & -2147.7 & -2154.55 \\
PCF-PK1 & $E(0^+)$ & -2114.01 & -2123.8 & -2132 & -2139.89 & -2147.49 & -2154.84 \\
        &$B_f-E(0^+)$  & 0.08 & -0.57 & -0.50 & -0.42 & -0.21 & 0.29
\vspace{0.2cm}\\
        &  $B_f $& -2118.88 & -2127.26 & -2135.86 & -2143.53 & -2150.87 & -2157.93 \\
DD-PC1  & $E(0^+)$ & -2118.45 & -2126.88 & -2136.20 & -2144.17 & -2151.87 & -2159.23 \\
        & $B_f-E(0^+)$ & -0.43 & -0.38 & 0.34 & 0.64 & 1.00 & 1.23\\
\hline\hline
 \label{Ds-data}
\end{tabular}
\end{center}
\end{table*}

\setcounter{figure}{0}
\renewcommand{\thefigure}{S\arabic{figure}}

\begin{figure*}[b]
        \centering{\includegraphics[width=0.9\textwidth]{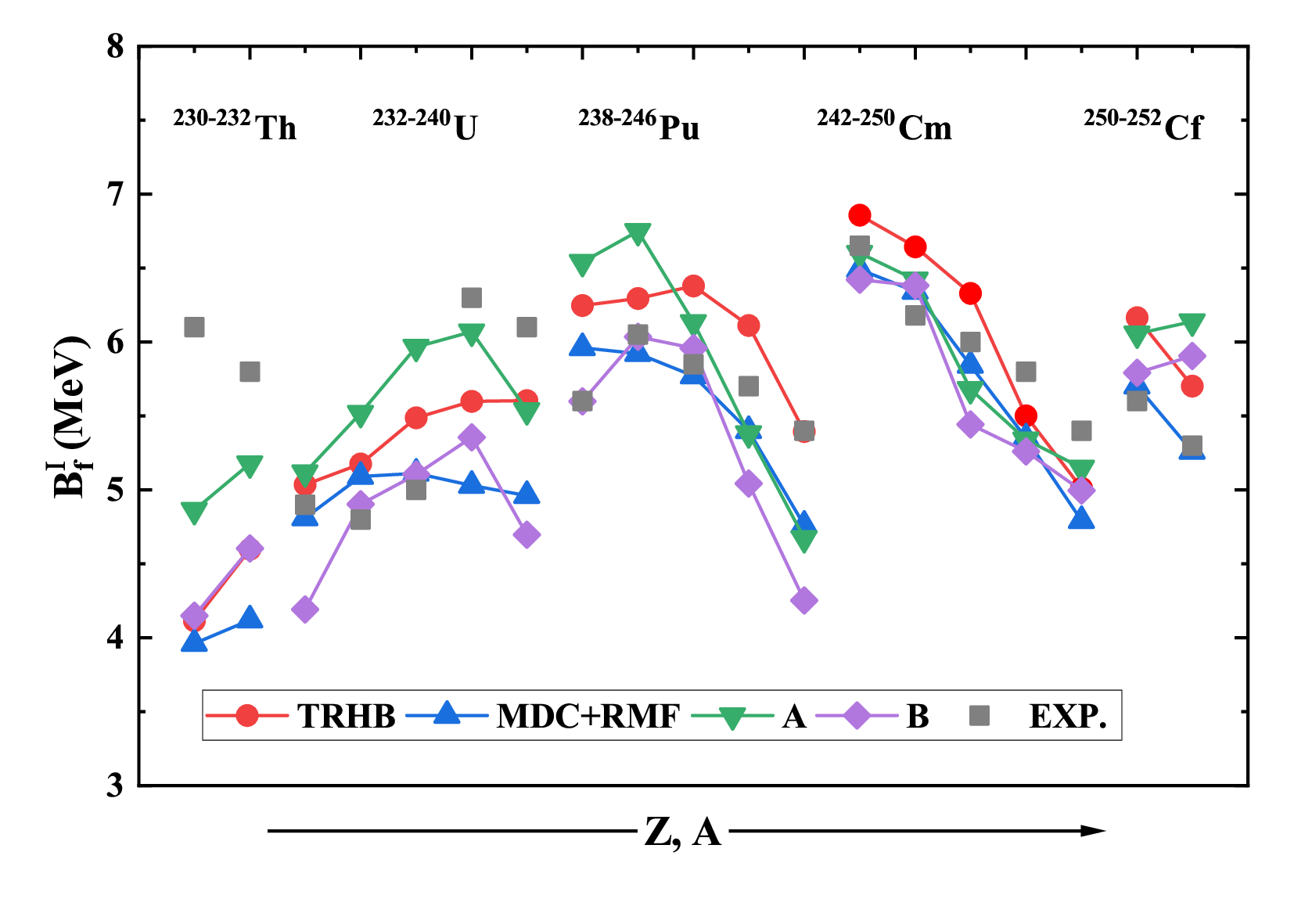}}
        \caption{\label{Bf}(Color online)
        The inner fission saddle heights of even-even actinide nuclei. Our results (TRHB, denoted by circles) are compared with those from MDC+RMF (triangles) \cite{Lu2014PRC} and HFB (diamonds) \cite{Delaroche2006NPA}, and empirical values (squares) \cite{CapoteNDS2009}.
       }
        \end{figure*}
    
\begin{figure*}[b]
\centering{\includegraphics[width=0.99\textwidth]{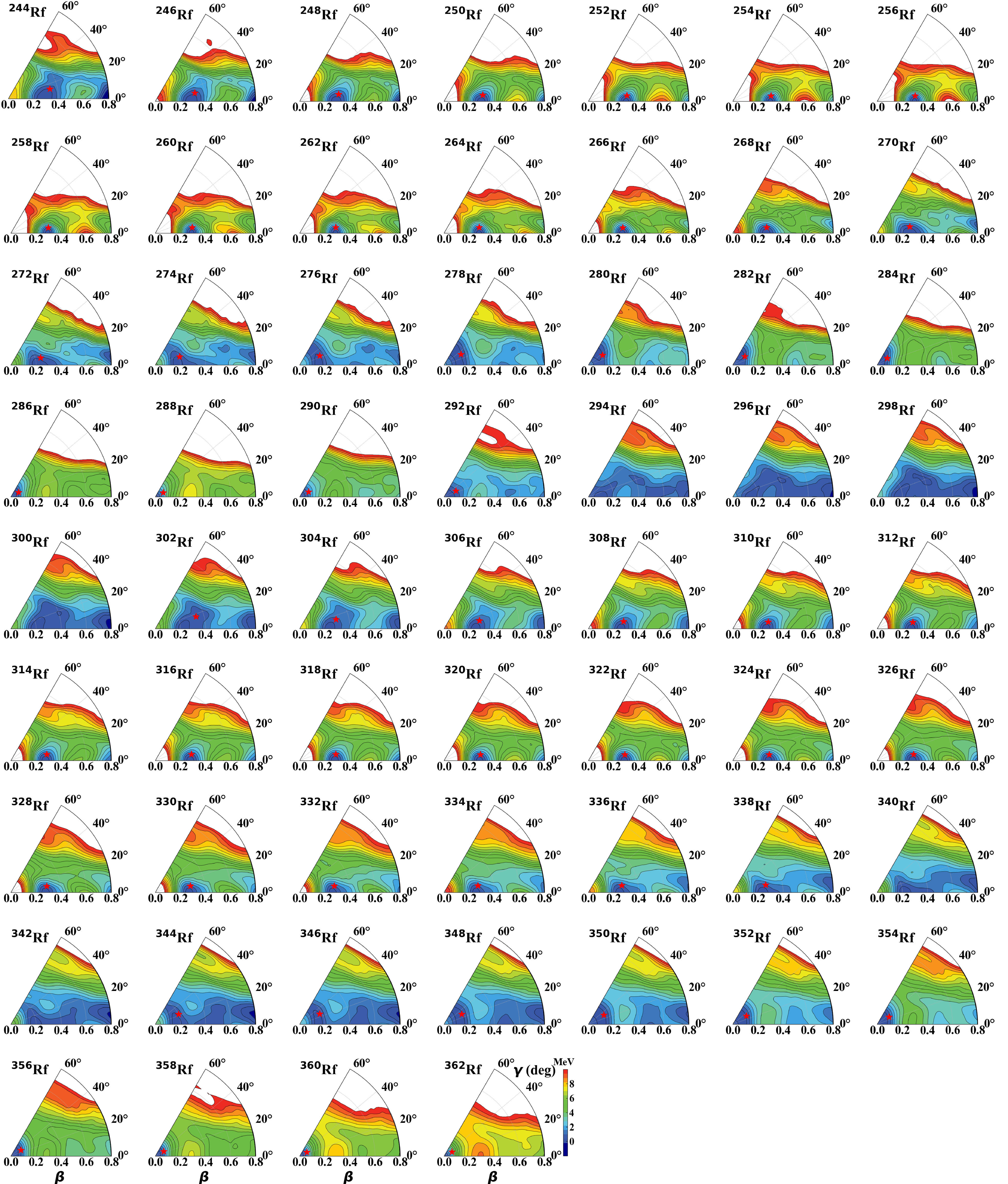}}
\caption{\label{Etot-Rf}(Color online) Collective potentials in the $(\beta, \gamma)$ plane  for the even-even Rf isotopic chain calculated by the constrained TRHB  with PC-PK1 functional \cite{Zhao2010PRC}. All energies are normalized with respect to the energy of the lowest minimum within the fission saddle. The energy difference between adjacent contour lines is 0.5 MeV. The red stars indicate the average deformations of the ground states $0^+_1$.}
\end{figure*}

\begin{figure*}[ht]
\centering{\includegraphics[width=0.99\textwidth]{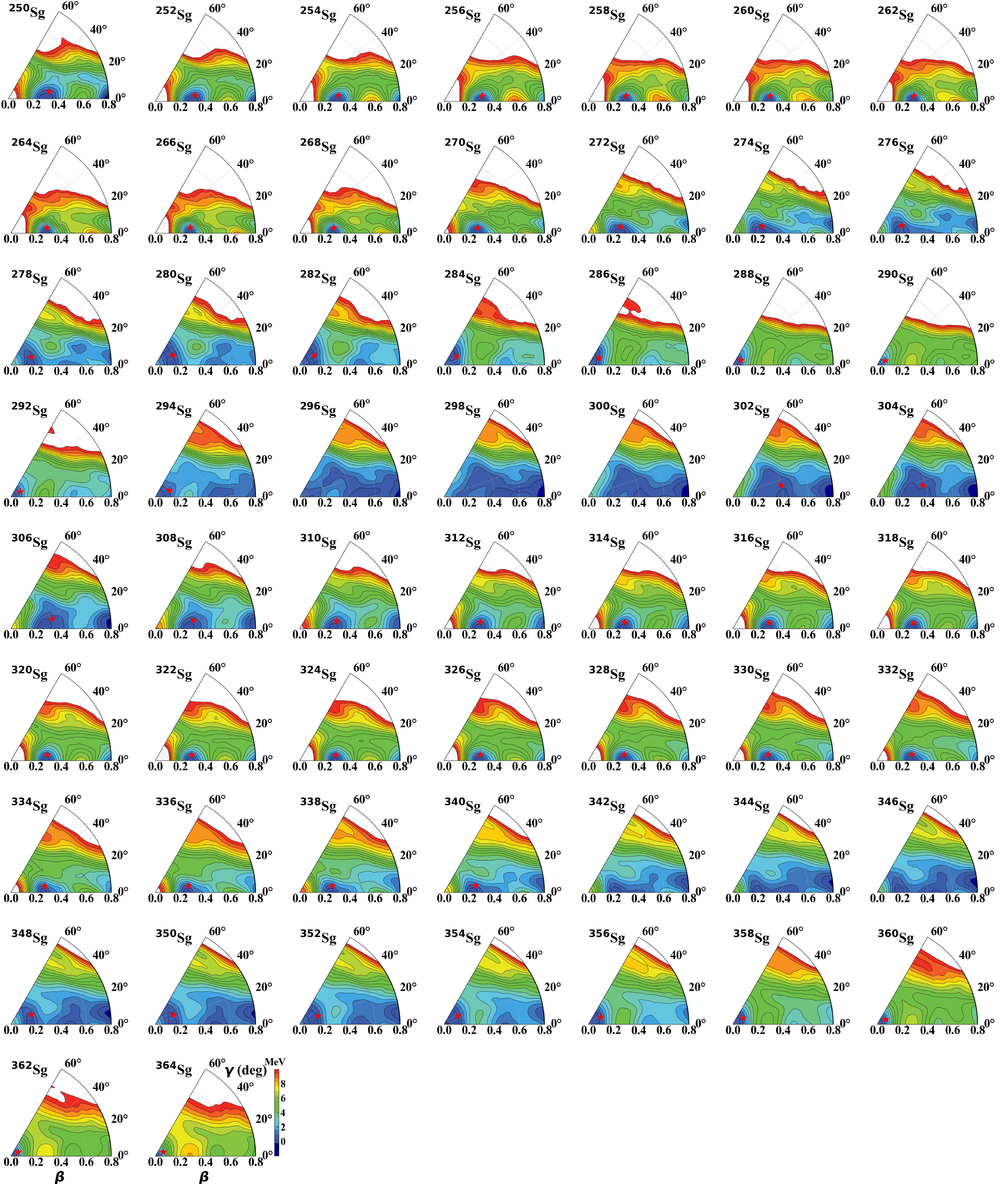}}
\caption{\label{Etot-Sg}(Color online) Same as Fig. \ref{Etot-Rf}, but for Sg isotopic chain.}
\end{figure*}

\begin{figure*}[ht]
\centering{\includegraphics[width=0.99\textwidth]{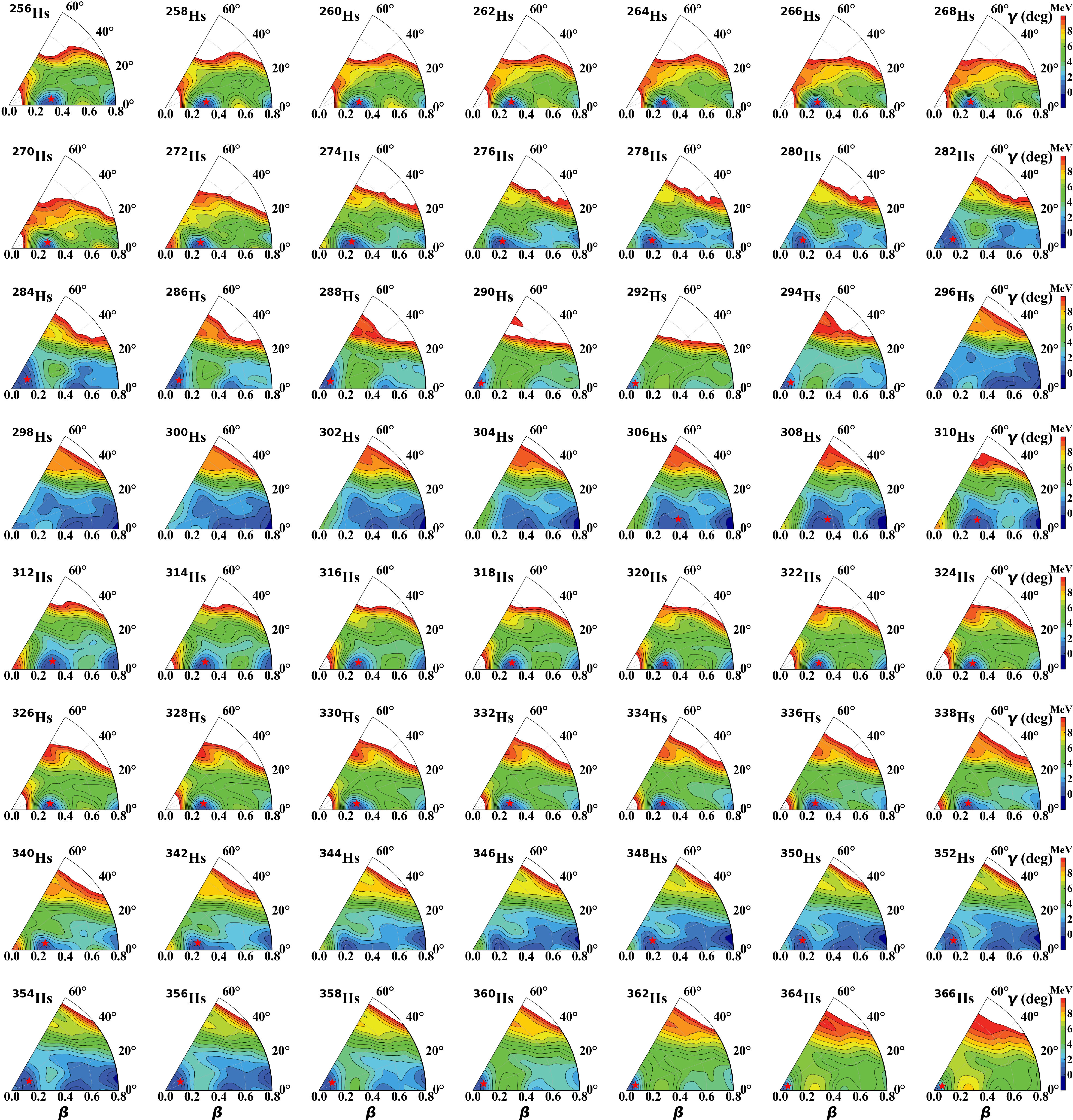}}
\caption{\label{Etot-Hs}(Color online) Same as Fig. \ref{Etot-Rf}, but for Hs isotopic chain.}
\end{figure*}

\begin{figure*}[ht]
\centering{\includegraphics[width=0.99\textwidth]{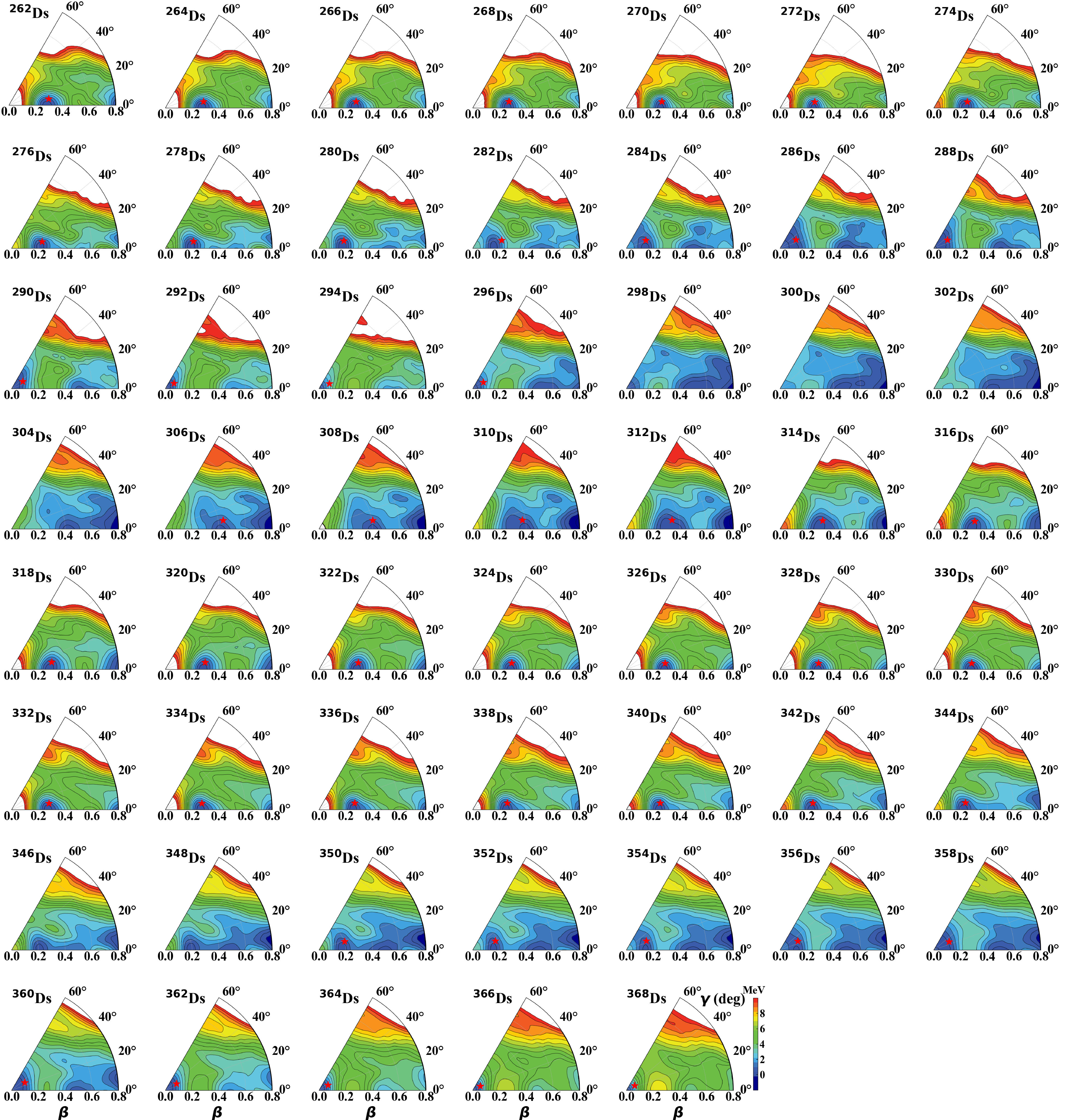}}
\caption{\label{Etot-Ds}(Color online) Same as Fig. \ref{Etot-Rf}, but for Ds isotopic chain.}
\end{figure*}

\begin{figure*}[ht]
\centering{\includegraphics[width=0.99\textwidth]{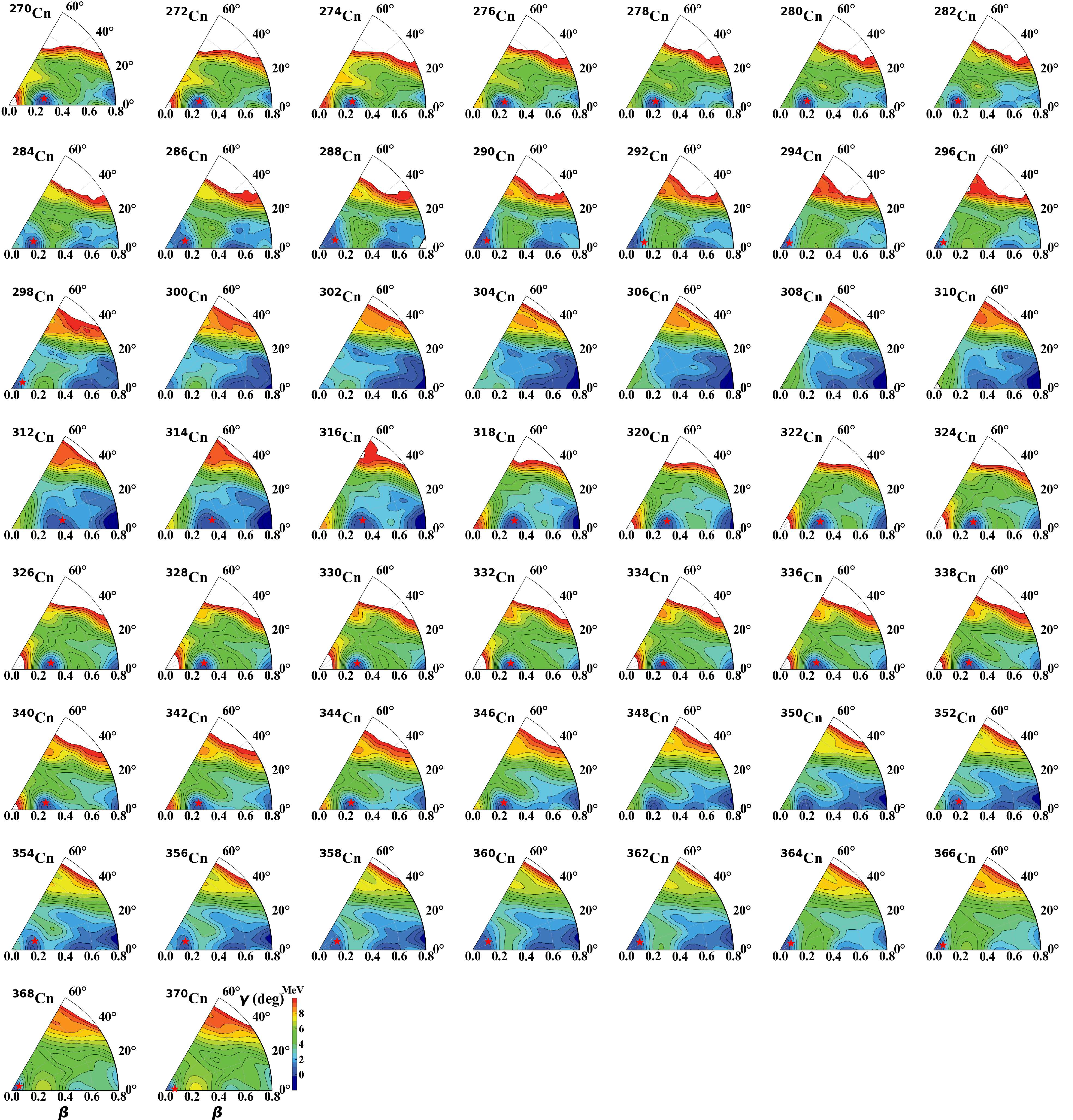}}
\caption{\label{Etot-Cn}(Color online) Same as Fig. \ref{Etot-Rf}, but for Cn isotopic chain.}
\end{figure*}

\begin{figure*}[ht]
\centering{\includegraphics[width=0.99\textwidth]{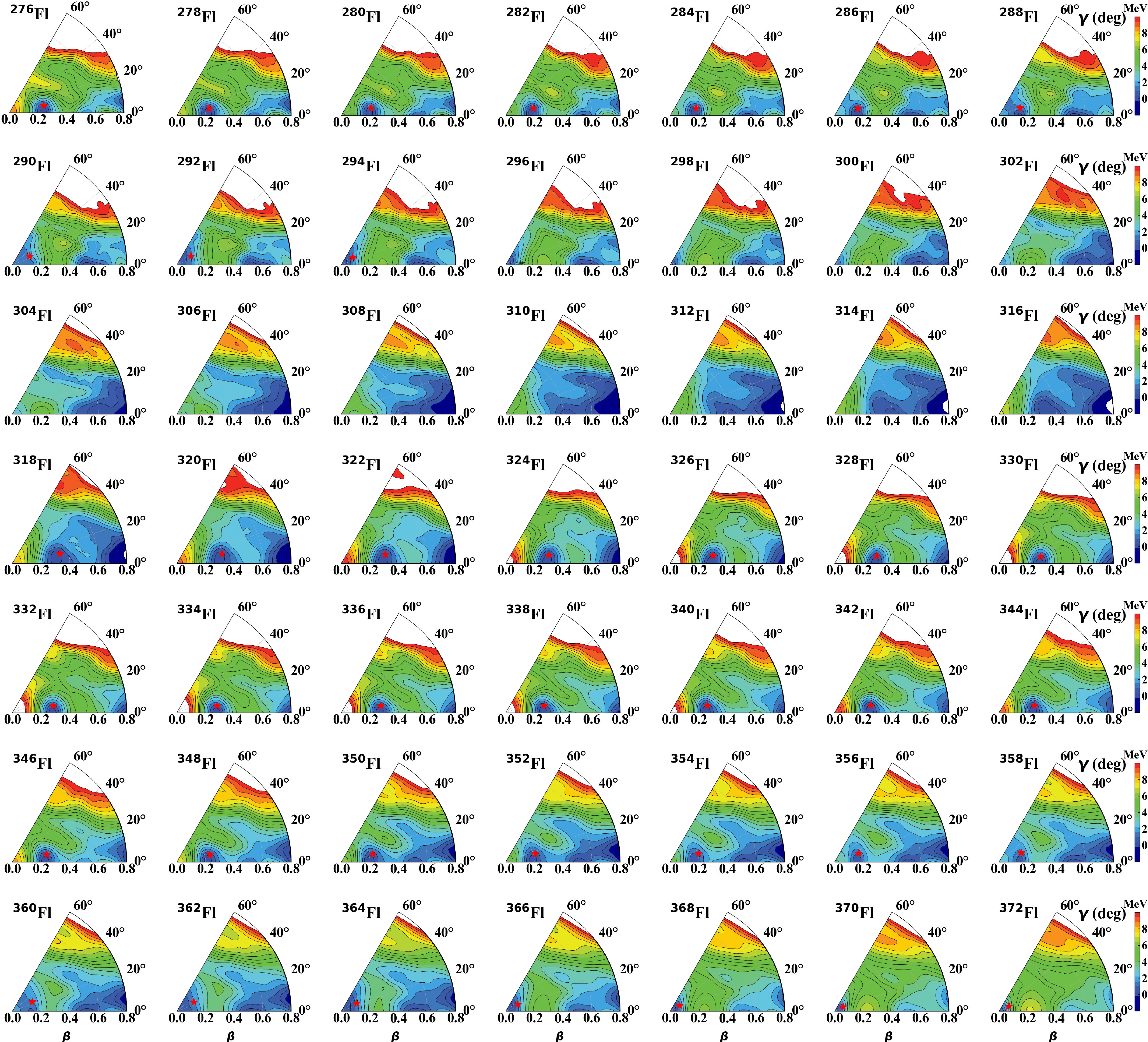}}
\caption{\label{Etot-Fl}(Color online) Same as Fig. \ref{Etot-Rf}, but for Fl isotopic chain.}
\end{figure*}

\begin{figure*}[ht]
\centering{\includegraphics[width=0.99\textwidth]{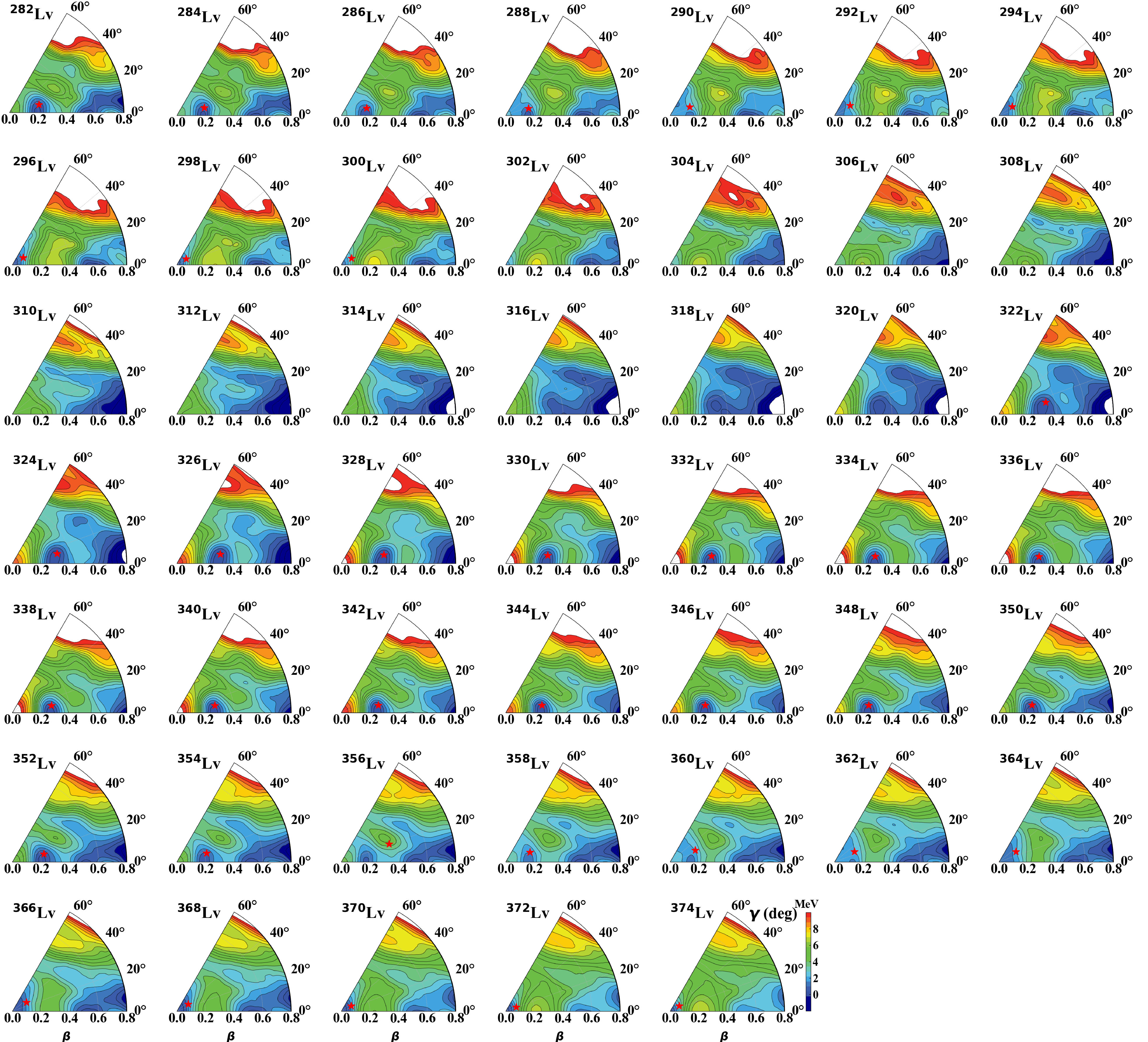}}
\caption{\label{Etot-Lv}(Color online) Same as Fig. \ref{Etot-Rf}, but for Lv isotopic chain.}
\end{figure*}

\begin{figure*}[ht]
\centering{\includegraphics[width=0.99\textwidth]{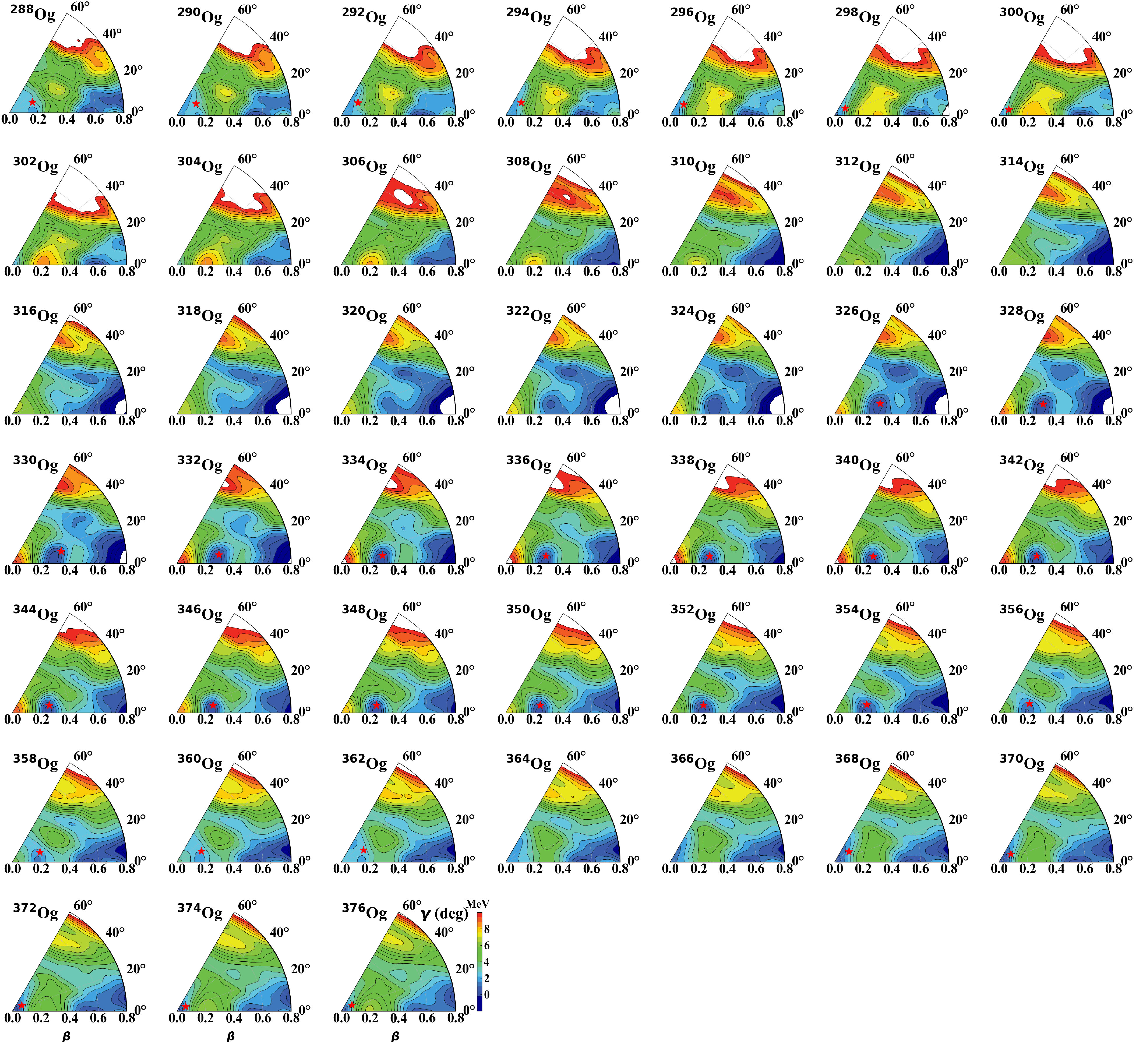}}
\caption{\label{Etot-Og}(Color online) Same as Fig. \ref{Etot-Rf}, but for Og isotopic chain.}
\end{figure*}

\begin{figure*}[ht]
\centering{\includegraphics[width=0.99\textwidth]{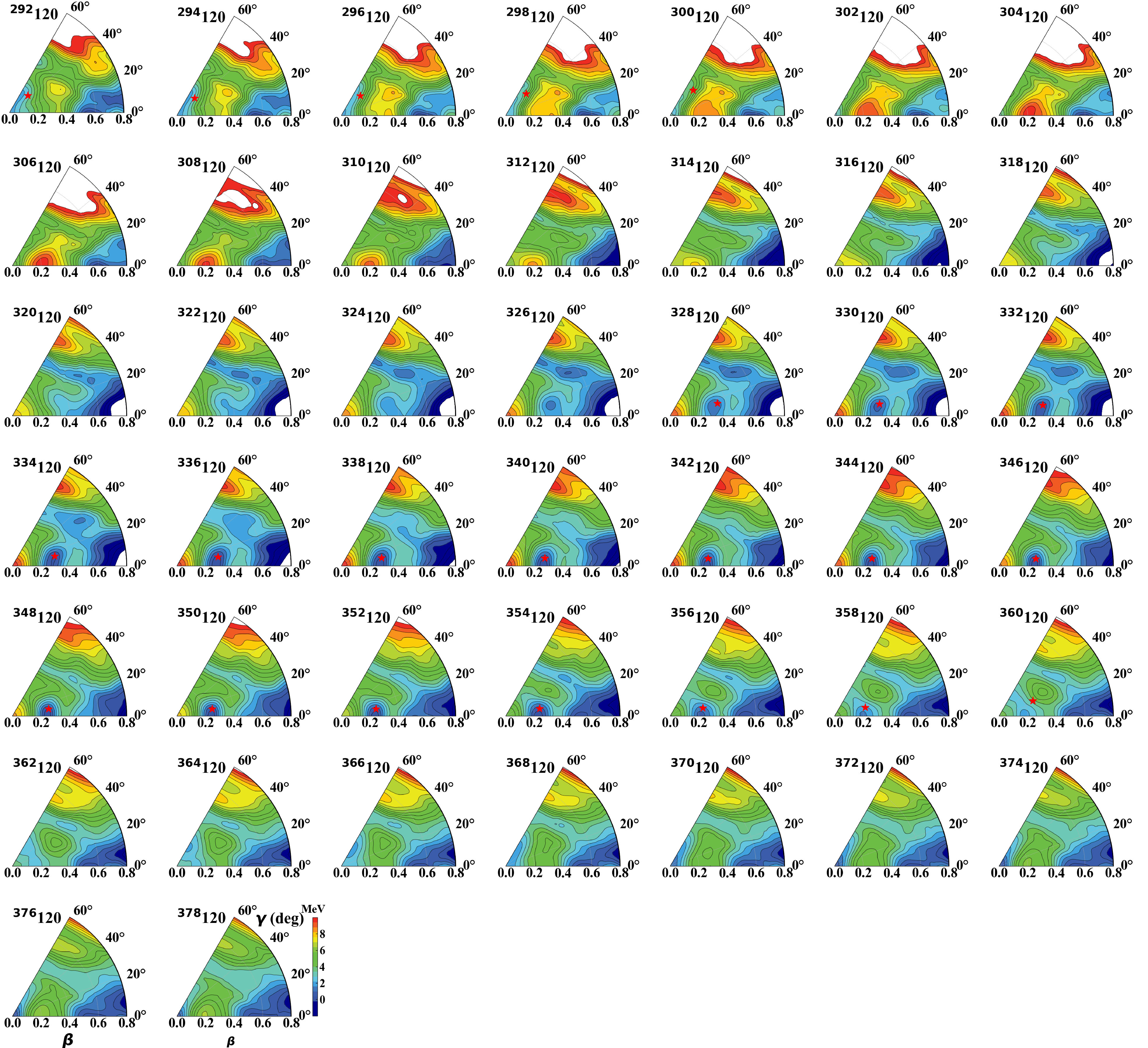}}
\caption{\label{Etot-120}(Color online) Same as Fig. \ref{Etot-Rf}, but for Z=120 isotopic chain.}
\end{figure*}

\begin{figure*}[ht]
\centering{\includegraphics[width=0.99\textwidth]{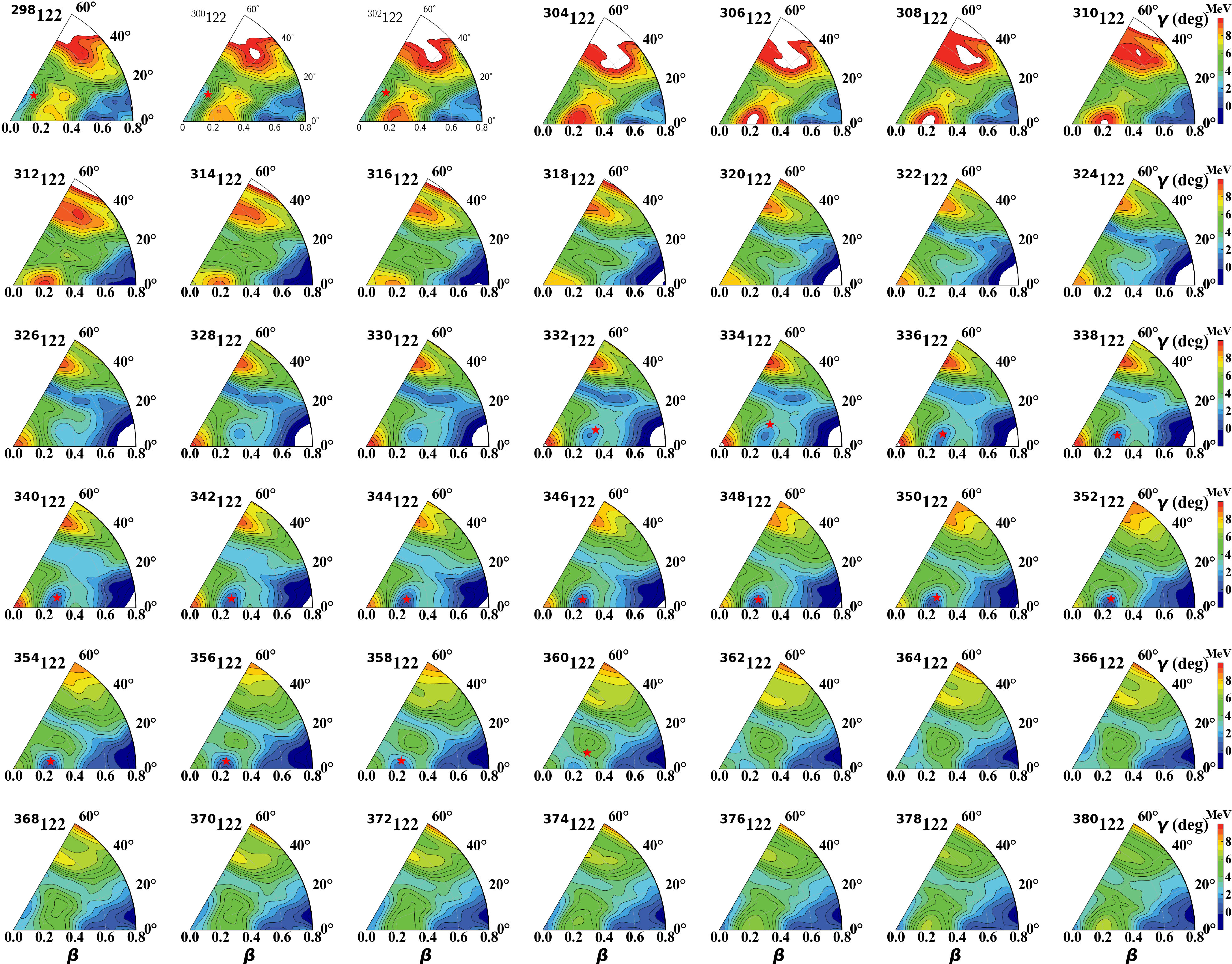}}
\caption{\label{Etot-122}(Color online) Same as Fig. \ref{Etot-Rf}, but for Z=122 isotopic chain.}
\end{figure*}

\begin{figure*}[ht]
\centering{\includegraphics[width=0.99\textwidth]{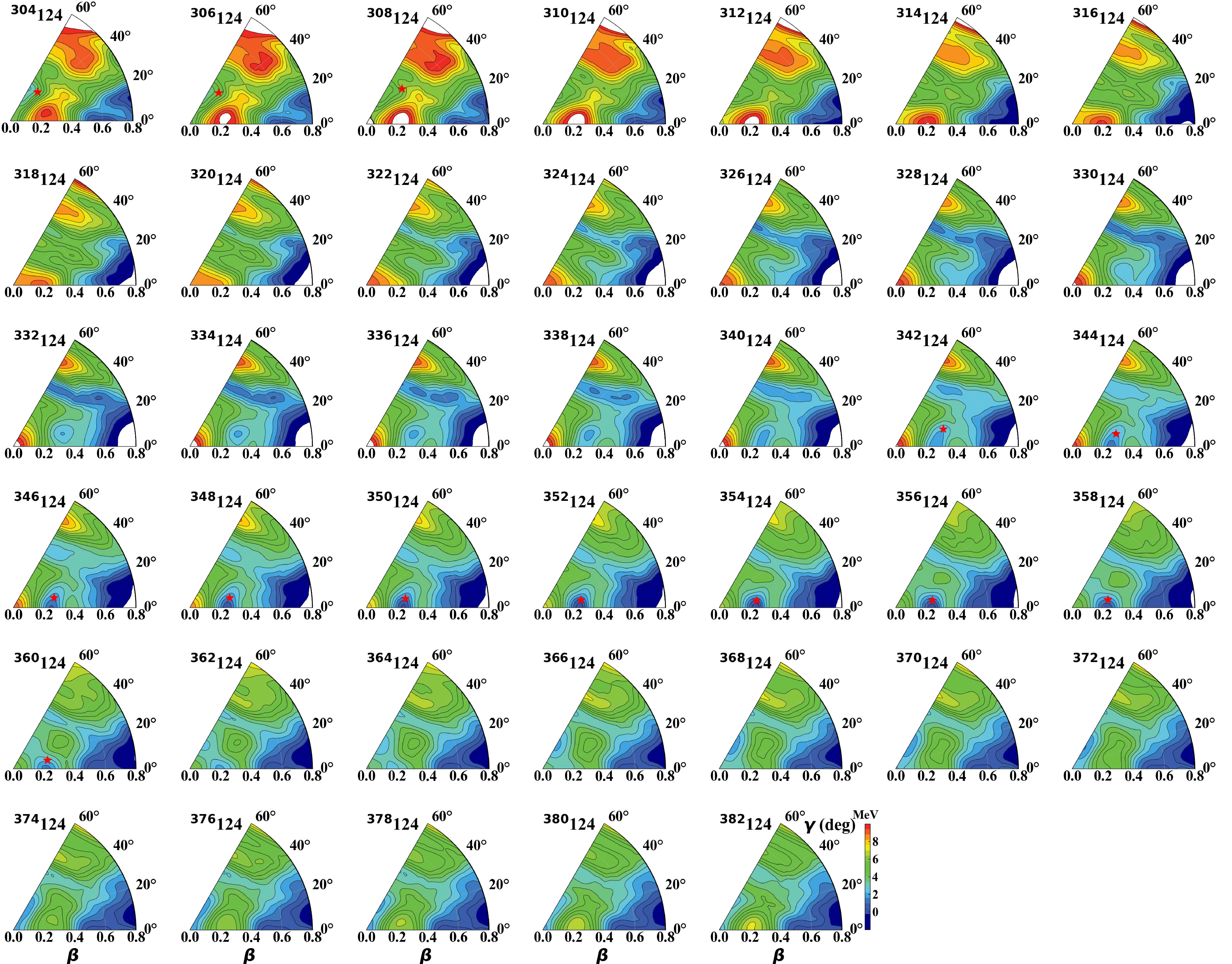}}
\caption{\label{Etot-124}(Color online) Same as Fig. \ref{Etot-Rf}, but for Z=124 isotopic chain.}
\end{figure*}

\begin{figure*}[ht]
\centering{\includegraphics[width=0.99\textwidth]{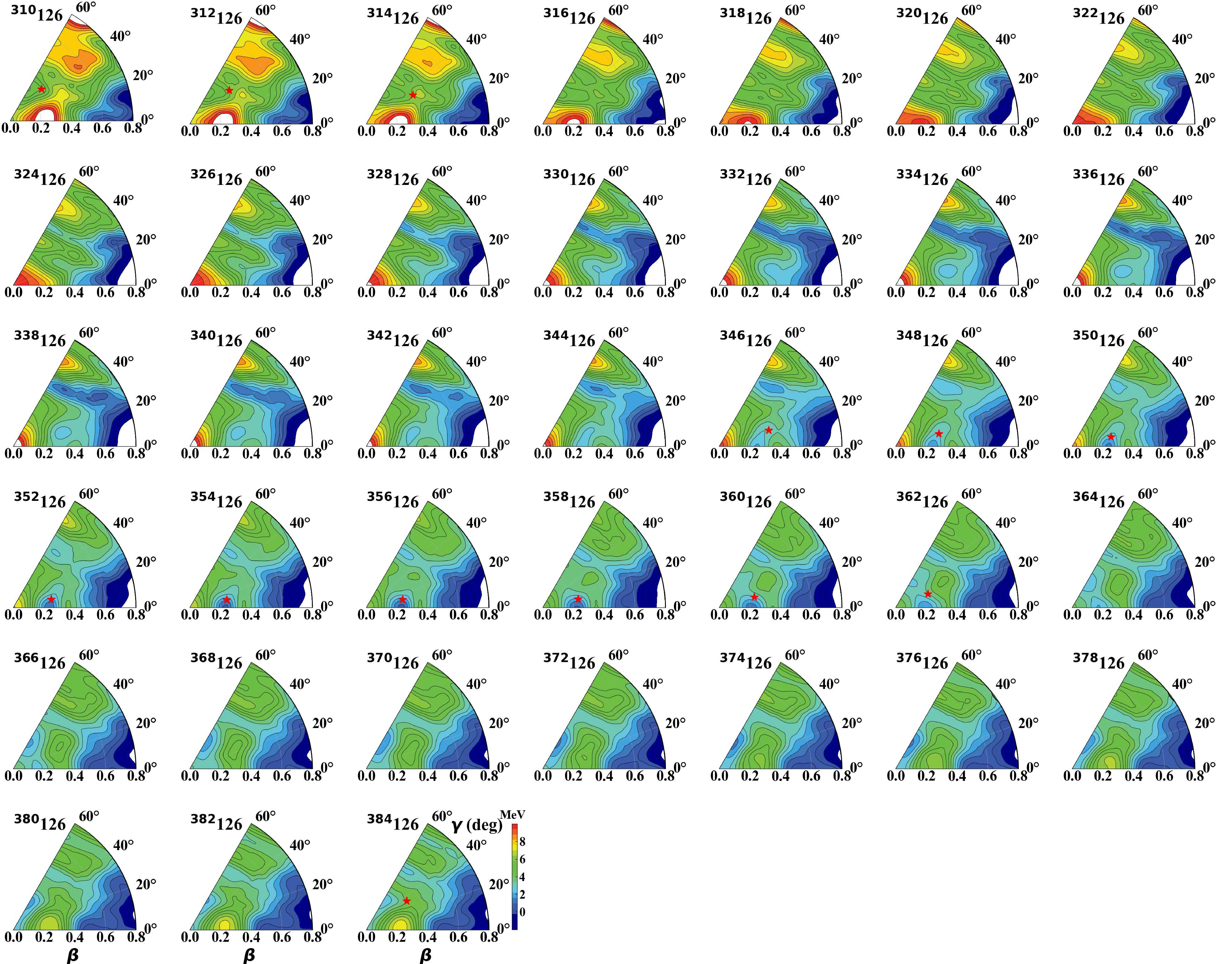}}
\caption{\label{Etot-126}(Color online) Same as Fig.\ref{Etot-Rf}, but for Z=126 isotopic chain.}
\end{figure*}

\begin{figure*}[ht]
\centering{\includegraphics[width=0.96\textwidth]{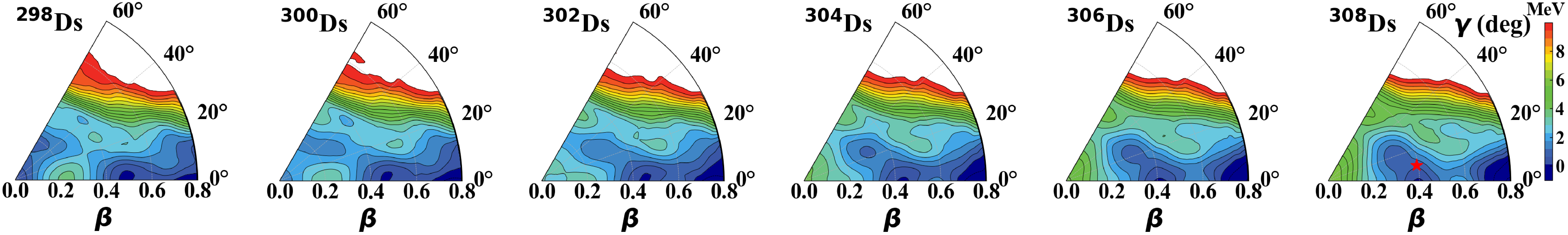}}
\caption{\label{PCF-PK1-Ds}(Color online) Collective potentials in the $(\beta, \gamma)$ plane  for some selected Ds isotopes calculated by the constrained TRHB  with PCF-PK1 functional \cite{Zhao2022PRC}. All energies are normalized with respect to the energy of the lowest minimum within the fission saddle. The energy difference between adjacent contour lines is 0.5 MeV. The red stars indicate the average deformations of the ground states $0^+_1$.}
\end{figure*}

\begin{figure*}[ht]
\centering{\includegraphics[width=0.96\textwidth]{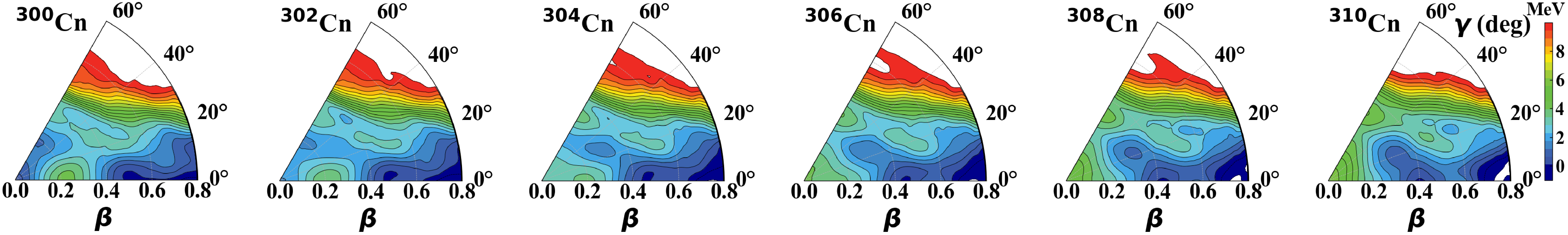}}
\caption{\label{PCF-PK1-Cn}(Color online) Same as Fig. \ref{PCF-PK1-Ds}, but for Cn isotopes.}
\end{figure*}

\begin{figure*}[ht]
\centering{\includegraphics[width=0.96\textwidth]{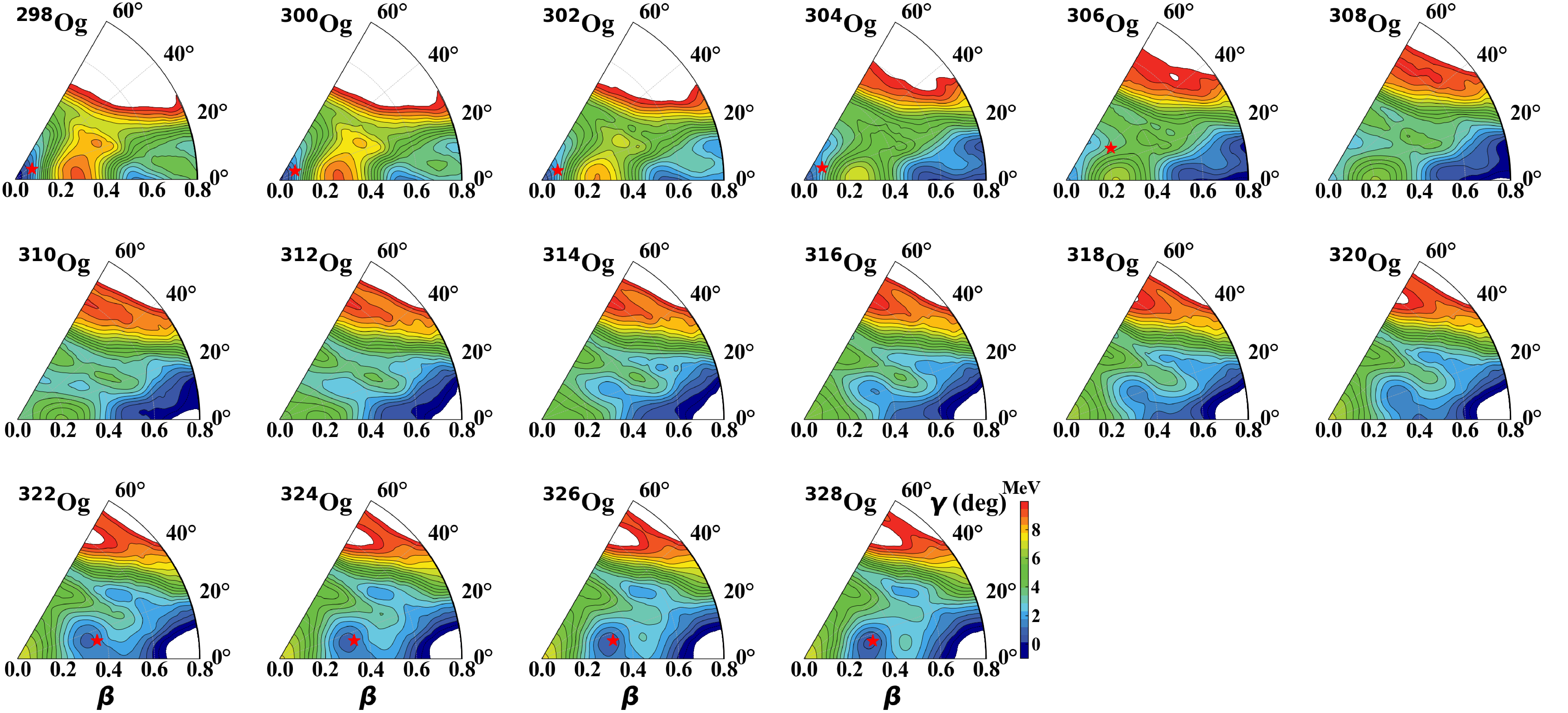}}
\caption{\label{PCF-PK1-Og}(Color online) Same as Fig. \ref{PCF-PK1-Ds}, but for Og isotopes.}
\end{figure*}

\begin{figure*}[ht]
\centering{\includegraphics[width=0.96\textwidth]{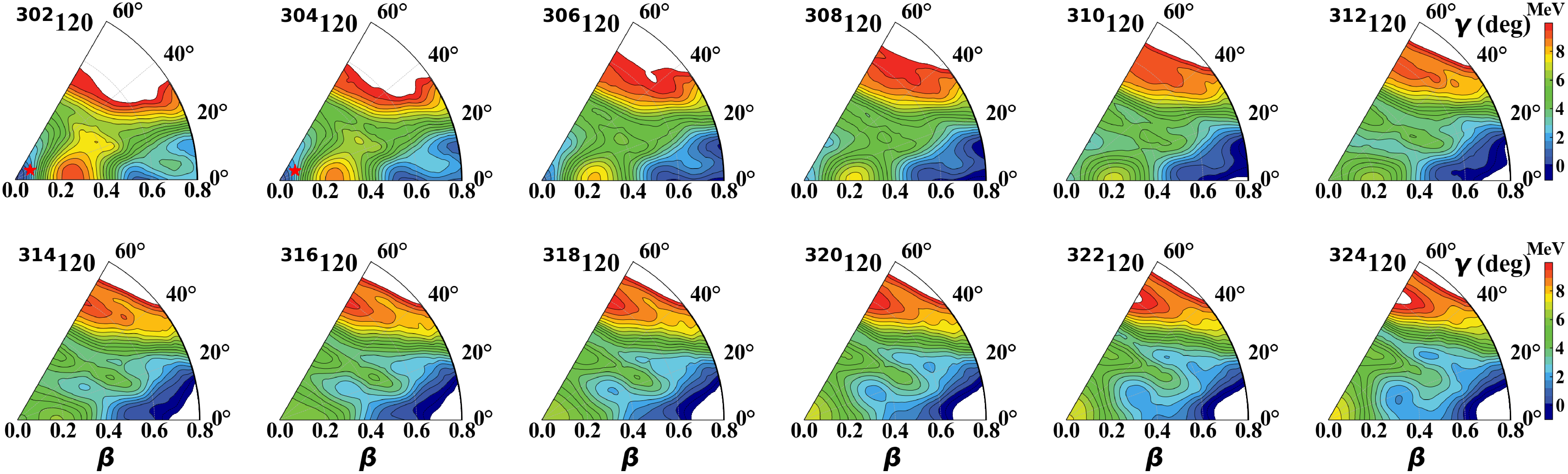}}
\caption{\label{PCF-PK1-20}(Color online) Same as Fig. \ref{PCF-PK1-Ds}, but for Z=120 isotopes.}
\end{figure*}

\begin{figure*}[ht]
\centering{\includegraphics[width=0.96\textwidth]{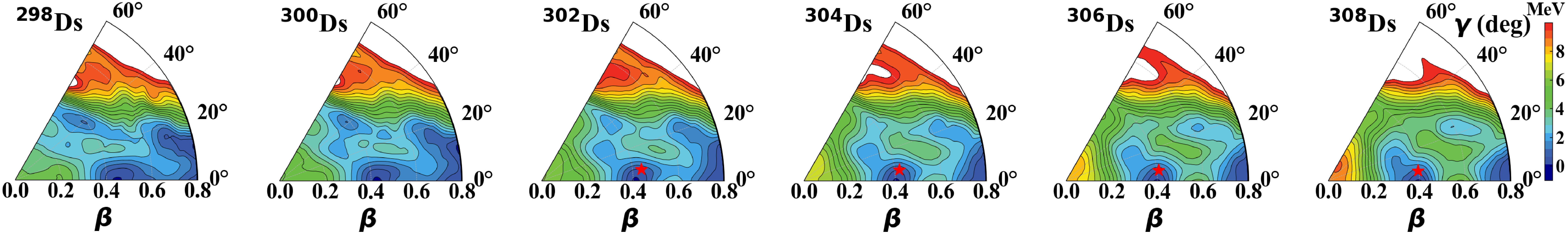}}
\caption{\label{DD-PC1-Ds}(Color online) Collective potentials in the $(\beta, \gamma)$ plane  for for some selected Ds isotopes calculated by the constrained TRHB  with DD-PC1 functional \cite{Niksic2008PRC}. All energies are normalized with respect to the energy of the lowest minimum within the fission saddle. The energy difference between adjacent contour lines is 0.5 MeV. The red stars indicate the average deformations of the ground states $0^+_1$.}
\end{figure*}

\begin{figure*}[ht]
\centering{\includegraphics[width=0.96\textwidth]{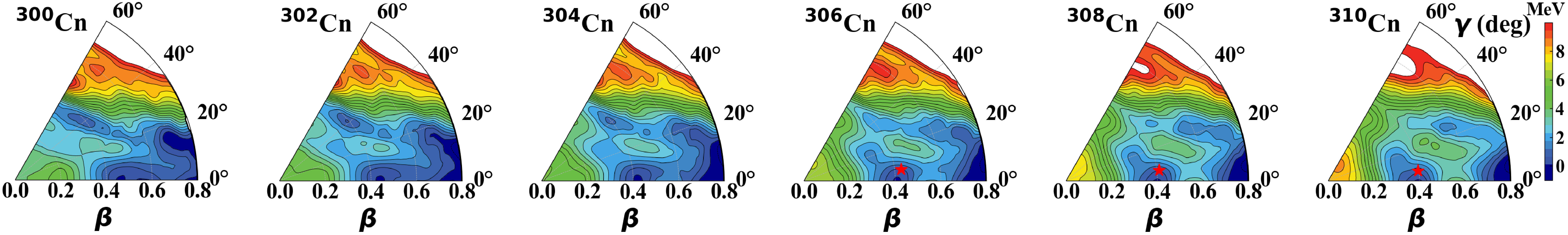}}
\caption{\label{DD-PC1-Cn}(Color online) Same as Fig. \ref{DD-PC1-Ds}, but for Cn isotopes.}
\end{figure*}

\begin{figure*}[ht]
\centering{\includegraphics[width=0.96\textwidth]{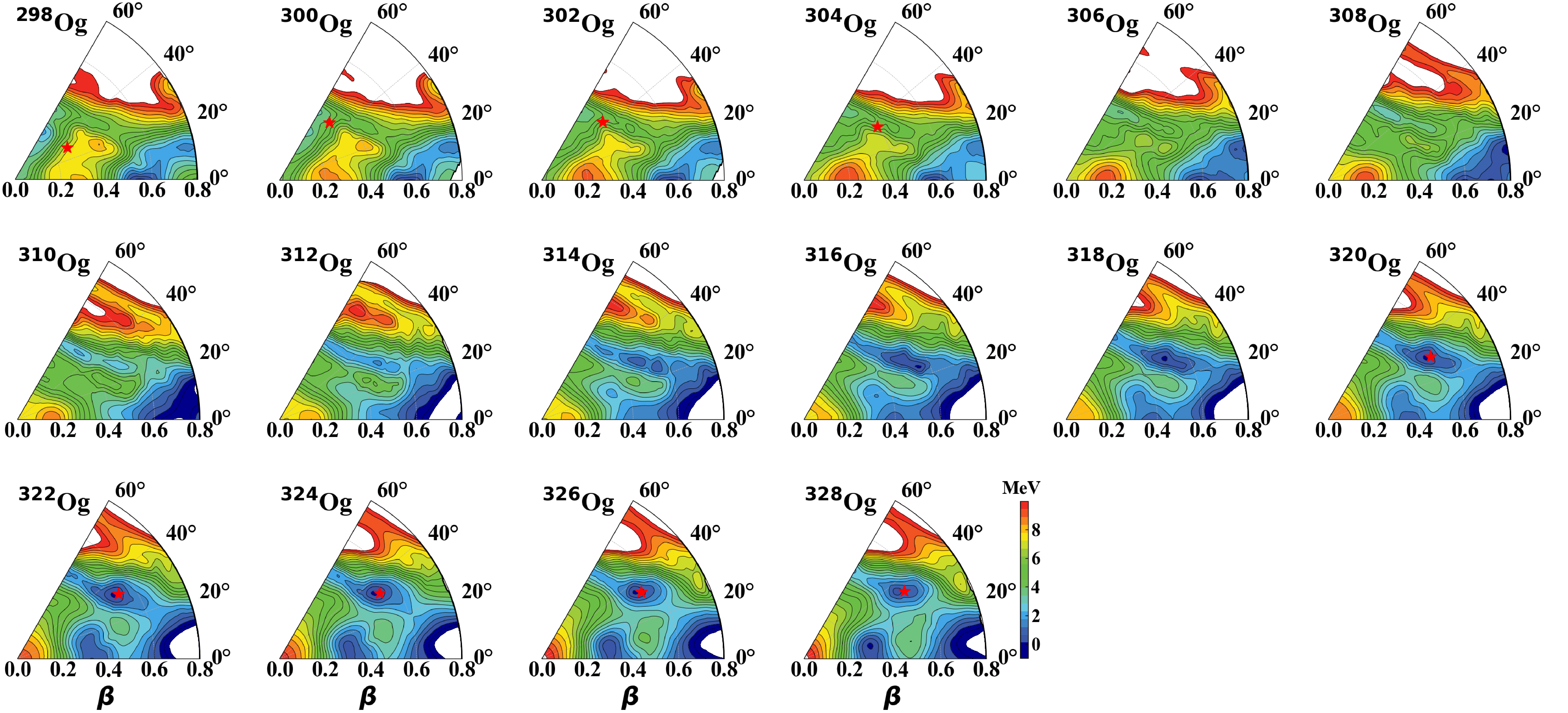}}
\caption{\label{DD-PC1-Og}(Color online) Same as Fig. \ref{DD-PC1-Ds}, but for Og isotopes.}
\end{figure*}

\begin{figure*}[ht]
\centering{\includegraphics[width=0.96\textwidth]{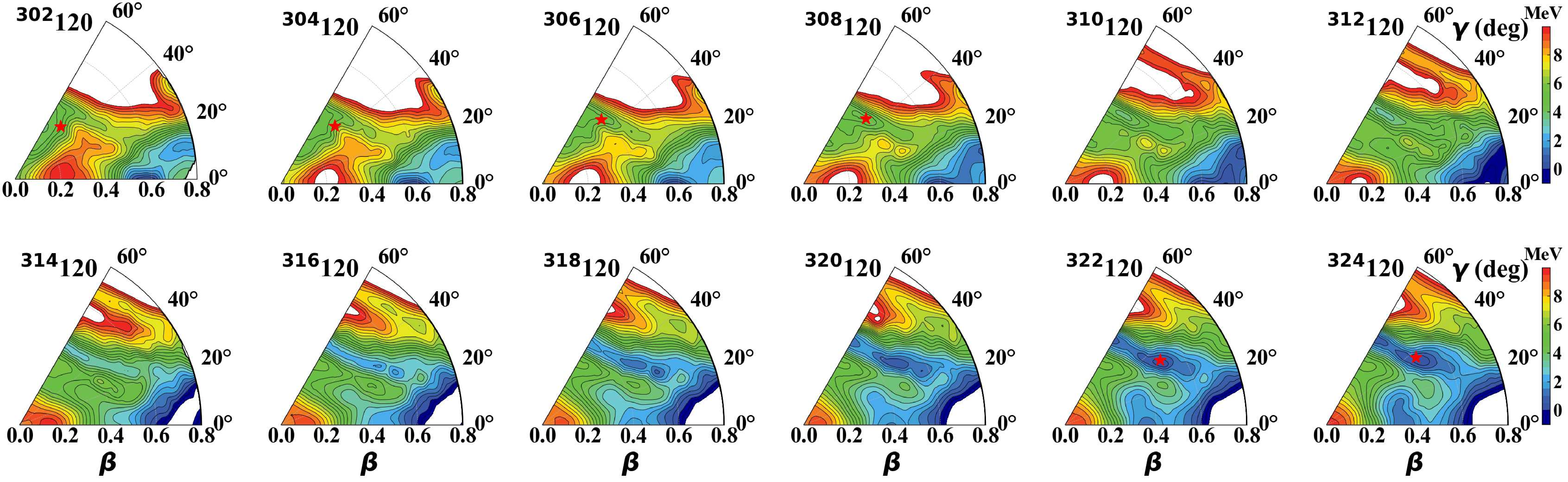}}
\caption{\label{DD-PC1-20}(Color online) Same as Fig. \ref{DD-PC1-Ds}, but for Z=120 isotopes.}
\end{figure*}

\begin{figure*}[ht]
    \centering{\includegraphics[width=0.96\textwidth]{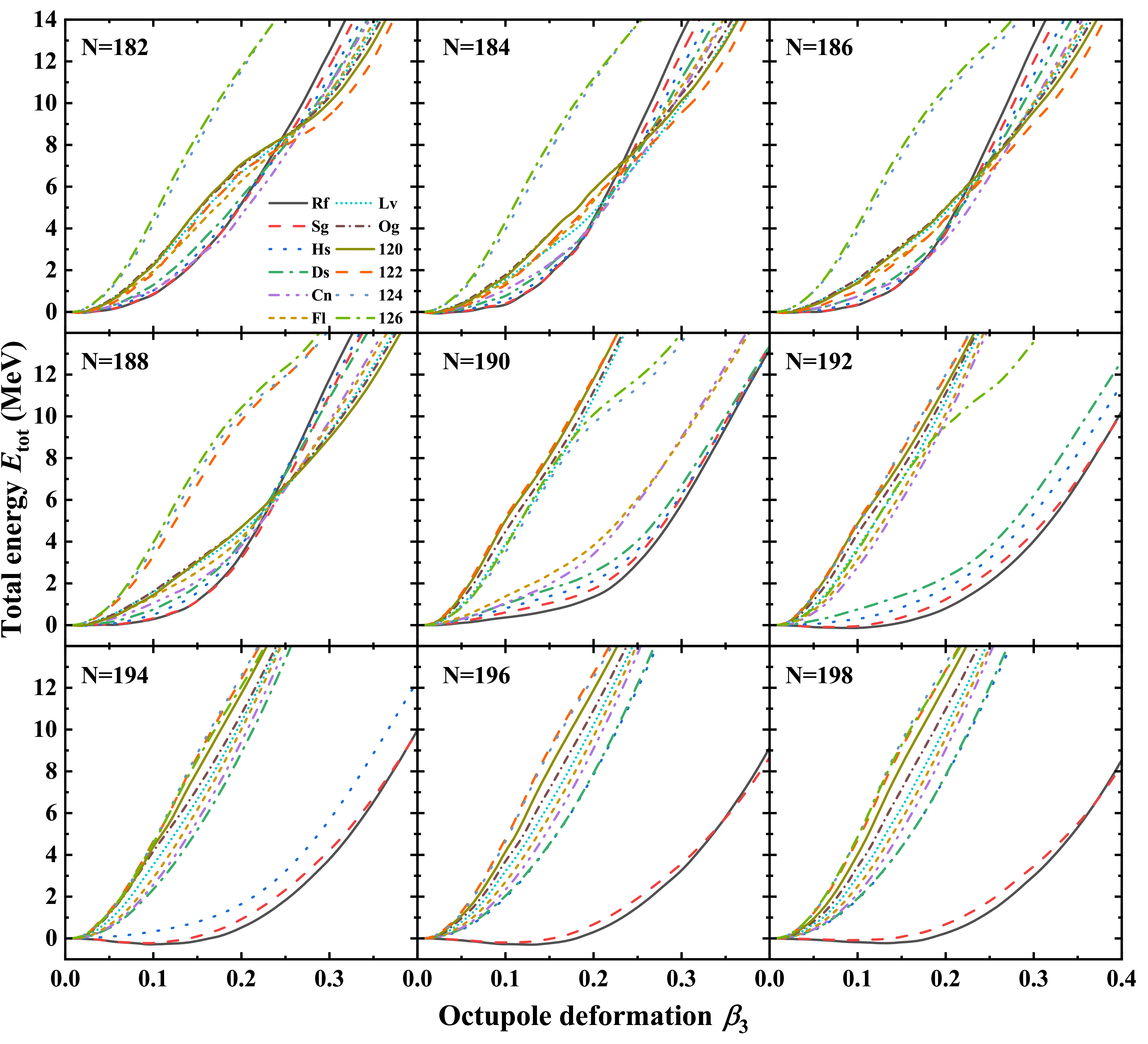}}
    \caption{\label{PEC_oct}(Color online)
    Potential energy curves (PECs) for the even-even $N=182-198 $ isotones as a function of the  octupole deformation  parameter $\beta_3$ calculated by the axially symmetric quadrupole-octupole relativistic Hartree-Bogoliubov model with PC-PK1 functional.  For clarity, the PECs are all normalized with respect to their corresponding quadrupole minima ($\beta_3=0$).}
    \end{figure*}

\begin{figure*}[htb]
\centering{\includegraphics[width=0.88\textwidth]{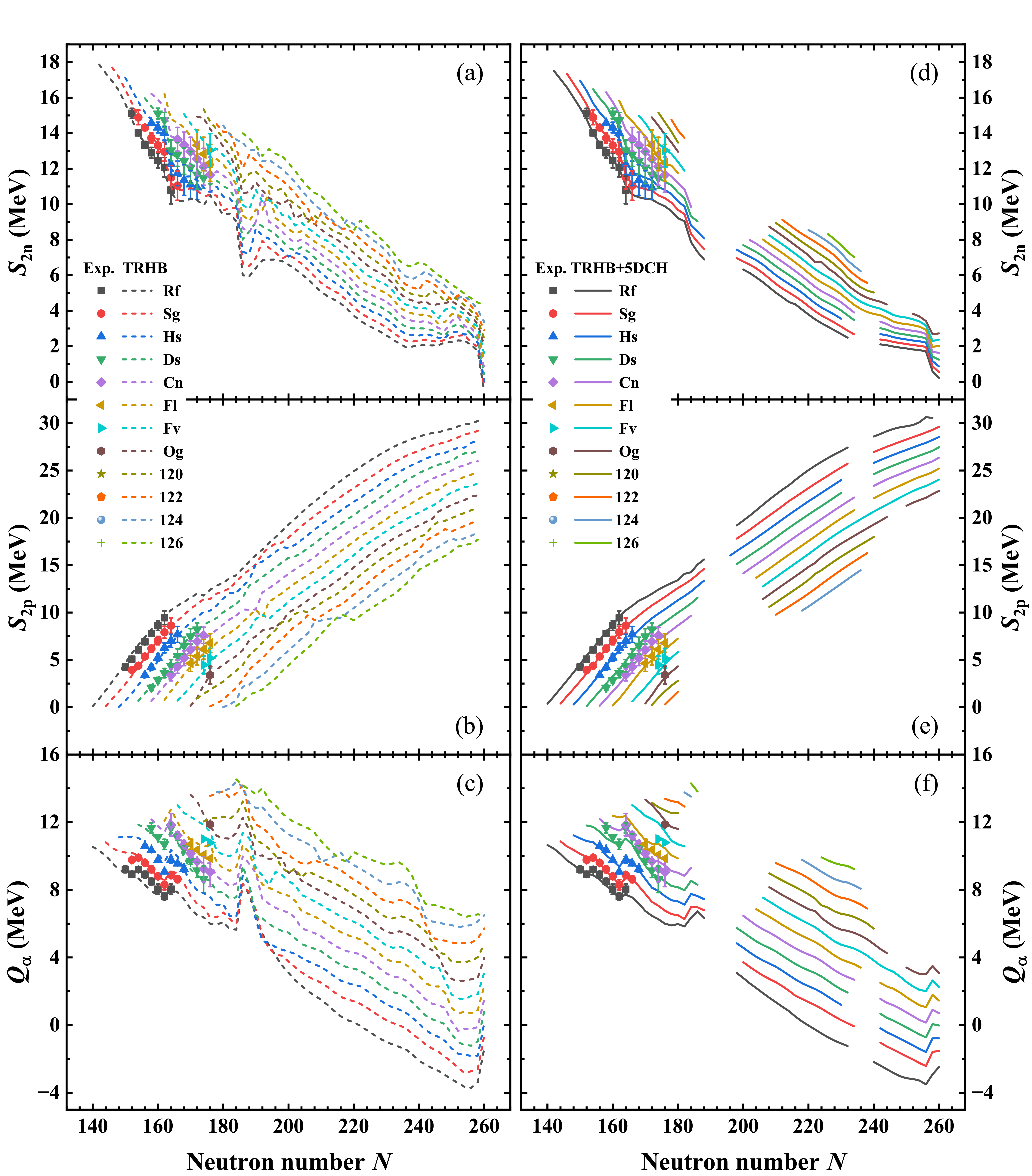}}
    \caption{(Color online) Two-neutron separation energies $S_{2n}$ (upper panels), two-proton separation energies $S_{2p}$ (middle panels), and $\alpha$ decay energies $Q_\alpha$ (lower panles) of SHN as functions of the neutron number calculated by the mean-field TRHB (dashed lines) and 5DCH (solid lines), in comparison with available data (solid symbols).}
    \label{S2n}
\end{figure*}

\begin{figure*}[htb]
        \centering{\includegraphics[width=0.88\textwidth]{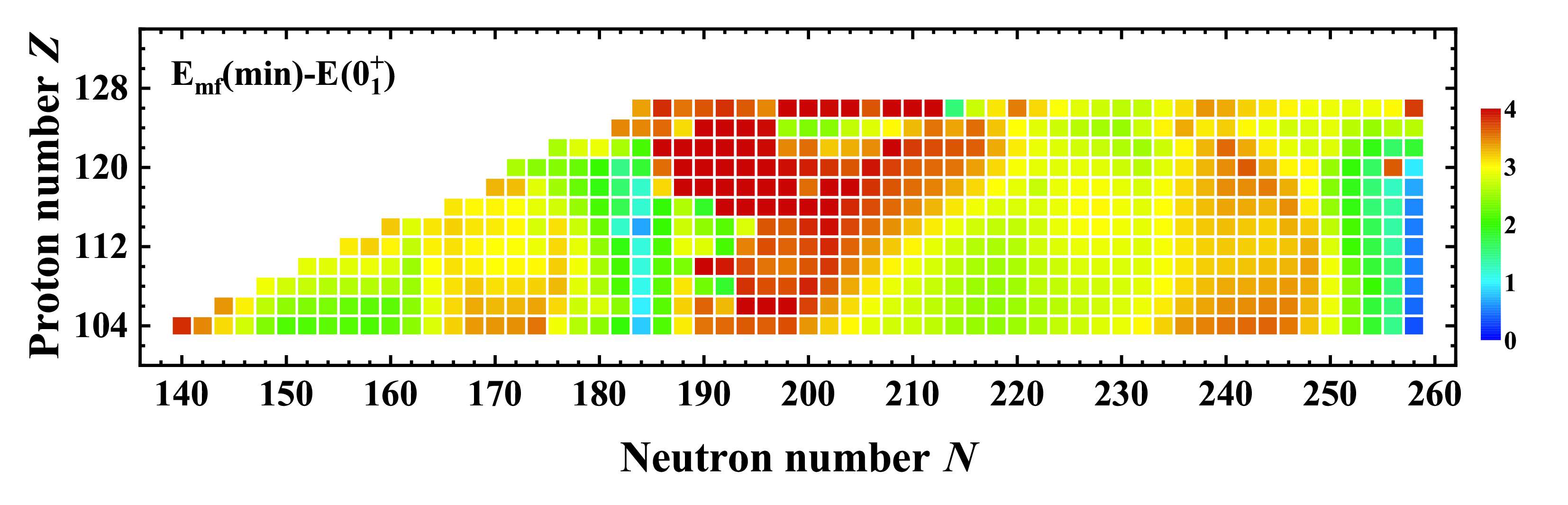}}
            \caption{(Color online) Dynamical correlation energies for all the calculated even-even superheavy nuclei.}
            \label{DCEs}
        \end{figure*}

\begin{figure*}[htb]
    \centering
    \includegraphics[width=0.88\linewidth]{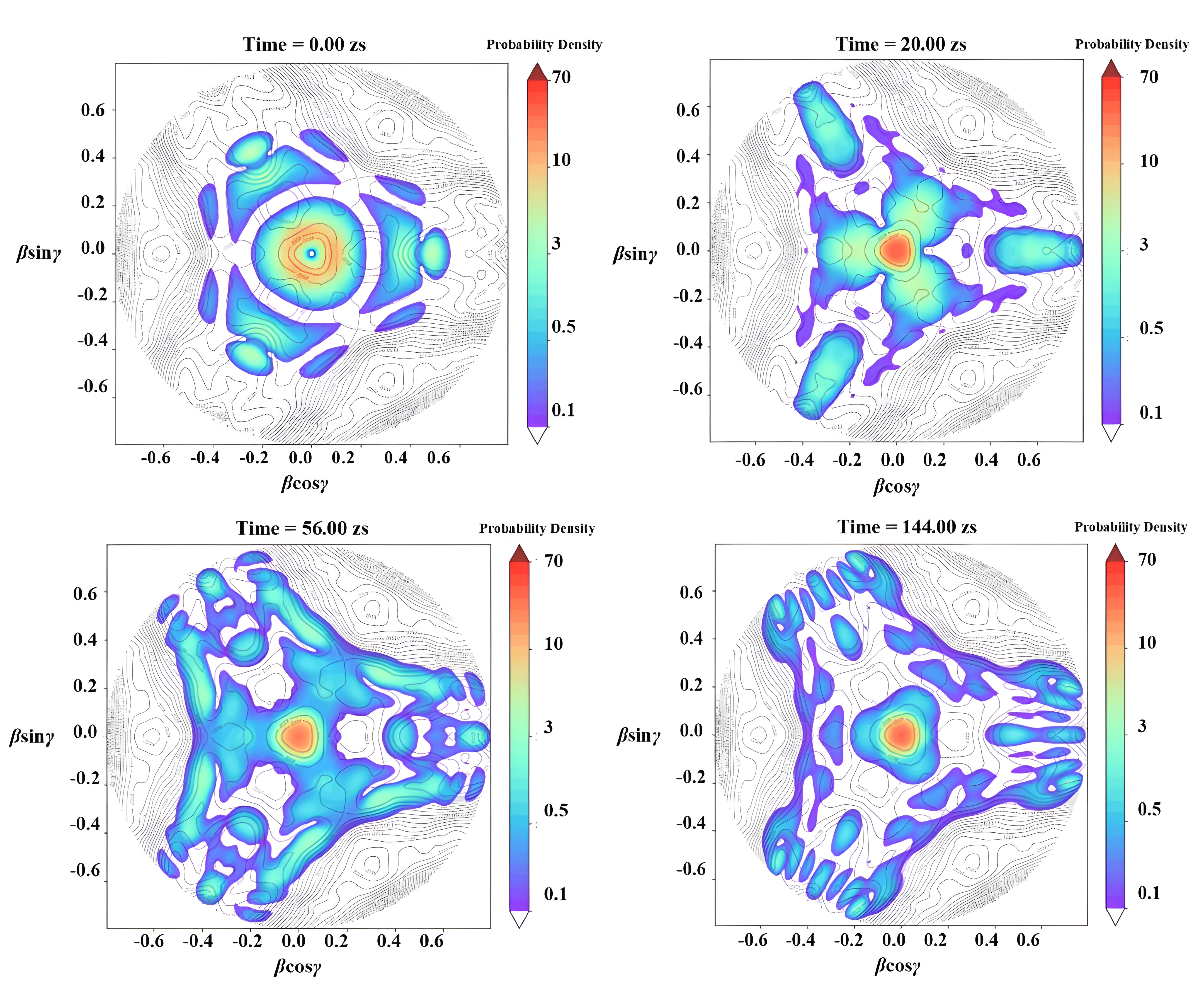}
    \caption{(Color online) Time evolution of the probability density distributions for the ``unbound'' state $0^+_1$ of $^{298}$Ds by solving the corresponding time-dependent Schr\"{o}dinger equation in the collective space. Four panels denote the results at $t=0, 20, 56$, and 144 zs ($10^{-21} s$), respectively. The gray contour lines denote the collective potential of $^{298}$Ds.}
    \label{fig1}
\end{figure*}

\end{document}